\documentclass[longauth]{aa}

\newcommand{\vincent}[1]{#1}

\usepackage{graphicx}
\usepackage[varg]{txfonts}
\usepackage[T1]{fontenc}
\usepackage{aas_macros}
\usepackage{amsmath}
\usepackage{amssymb}
\usepackage{url}
\usepackage[colorlinks=true,citecolor=blue]{hyperref}

\usepackage{natbib}
\usepackage{subeqnarray}
\usepackage{color}

\usepackage{etoolbox}
\makeatletter
\patchcmd\@combinedblfloats{\box\@outputbox}{\unvbox\@outputbox}{}{%
   \errmessage{\noexpand\@combinedblfloats could not be patched}%
}%
 \makeatother
 
\begin{document}

\def\mo{\ifmmode^{-1}\else$^{-1}$\fi}
\def\Planck{\textit{Planck}}
\def\deg{\ifmmode^\circ\else$^\circ$\fi}
\def\GHz{\ifmmode $\,GHz$\else \,GHz\fi}
\def\MJysr{\ifmmode \,$MJy\,sr\mo$\else \,MJy\,sr\mo\fi}

\newbox\tablebox    \newdimen\tablewidth
\def\leaderfil{\leaders\hbox to 5pt{\hss.\hss}\hfil}
%
%
\def\endPlancktable{\tablewidth=\columnwidth
    $$\hss\copy\tablebox\hss$$
    \vskip-\lastskip\vskip -2pt}
\def\endPlancktablewide{\tablewidth=\textwidth
    $$\hss\copy\tablebox\hss$$
    \vskip-\lastskip\vskip -2pt}
\def\tablenote#1 #2\par{\begingroup \parindent=0.8em
    \abovedisplayshortskip=0pt\belowdisplayshortskip=0pt
    \noindent
    $$\hss\vbox{\hsize\tablewidth \hangindent=\parindent \hangafter=1 \noindent
    \hbox to \parindent{$^#1$\hss}\strut#2\strut\par}\hss$$
    \endgroup}
\def\doubleline{\vskip 3pt\hrule \vskip 1.5pt \hrule \vskip 5pt}

\def\daymodel{30JAN2017}
\def\revisedmodel{25JUL}
\def\laterevisedmodel{4SEP}

%
\def\cf{\emph{cf.\/}}
\def\ie{\emph{i.e.\/}}
\def\etc{etc.}
\def\apriori{\emph{a priori\/}}
\def\aposteriori{\emph{a posteriori\/}}
\def\afortiori{\emph{a fortiori\/}}
\def\loccit{\emph{a loc. cit.\/}}
\def\etal{\emph{et al.\/}}
\def\vg{\emph{v.g.\/}}
\def\eg{\emph{e.g.\/}}
\def\e{{\rm e}}
\def\erf{{\rm erf}}
\def\kms{{\rm km\,s}$^{-1}$}
\def\cmc{{\rm cm}$^{-3}$}

\def\d{{\rm d}}

\def\Lyman{\emph{Lyman}}
\def\PaperI{Paper I}

\newcommand{\hi}{{\sc Hi}}

\def\T{\emph{(Top)}}
\def\B{\emph{(Bottom)}}
\def\L{\emph{(Left)}}
\def\M{\emph{(Middle)}}
\def\R{\emph{(Right)}}

\newlength{\thsize}
\setlength{\thsize}{0.67\hsize}
\newlength{\hhsize}
\setlength{\hhsize}{1.03\hsize}
\newlength{\khsize}
\setlength{\khsize}{0.93\hsize}
\newlength{\qhsize}
\setlength{\qhsize}{0.48\hsize}

\def\red{\emph{(red)}}
\def\blue{\emph{(blue)}}
\def\black{\emph{(black)}}
\def\pink{\emph{(pink)}}
\def\green{\emph{(green)}}
\def\lightblue{\emph{(light blue)}}
\def\darkblue{\emph{(dark blue)}}
\def\orange{\emph{(orange)}}
\def\mauve{\emph{(mauve)}}

\def\lmax{$\lambda_{\rm max}$}
\def\ebv{$E(B-V)$}

\def\Lper{L_{\perp}}
\def\Lpar{L_{\parallel}}
\def\Lparper{L_{\parallel,\perp}}
\def\Cparper{C_{\parallel,\perp}}
\def\Cpar{C_{\parallel}}
\def\Cper{C_{\perp}}

\def\NH{N_{\rm H}}
\def\cmsq{cm$^{-2}$}

\keywords{}
\def\Rv{R_{\rm V}}
\def\Av{A_{\rm V}}
\def\EBV{E(B-V)}

\def\symaxis{\hat{\bold{a}}}
\def\xaxis{\hat{\bold{x}}}
\def\yaxis{\hat{\bold{y}}}
\def\zaxis{\hat{\bold{z}}}

\def\Bpos{$\bold{B_{\rm pos}}$}
\def\Bmag{$\bold{B}$}

\def\plev{f_{\rm max}}
\def\pstiff{p_{\rm stiff}}
\def\athresh{a_{\rm alig}}
\def\CE{C^{\rm E}}
\def\CH{C^{\rm H}}
\def\CEext{C_{\rm ext}^{\rm E}}
\def\CHext{C_{\rm ext}^{\rm H}}
\def\CEabs{C_{\rm abs}^{\rm E}}
\def\CHabs{C_{\rm abs}^{\rm H}}
\def\CEsca{C_{\rm sca}^{\rm E}}
\def\CHsca{C_{\rm sca}^{\rm H}}

\def\Cextpara{C_{\rm 1,ext}}
\def\Cextperp{C_{\rm 2, ext}}
\def\Cabspara{C_{\rm 1, abs}}
\def\Cabsperp{C_{\rm 2, abs}}
\def\Cscapara{C_{\rm 1, sca}}
\def\Cscaperp{C_{\rm 2, sca}}

\def\Qextpara{Q_{\rm 1, ext}}
\def\Qextperp{Q_{\rm 2, ext}}
\def\Qabspara{Q_{\rm 1, abs}}
\def\Qabsperp{Q_{\rm 2, abs}}
\def\Qscapara{Q_{\rm 1, sca}}
\def\Qscaperp{Q_{\rm 2, sca}}

\def\Cpol{C_{\rm pol}}
\def\Cext{C_{\rm ext}}
\def\Cabs{C_{\rm abs}}
\def\Csca{C_{\rm sca}}
\def\Qpol{Q_{\rm pol}}
\def\Qext{Q_{\rm ext}}
\def\Qabs{Q_{\rm abs}}
\def\Qsca{Q_{\rm sca}}
\def\Cgeo{C^{\rm geo}}
\def\Cavg{C_{\rm <ext>}}
\def\Cavgabs{C_{\rm <abs>}}
\def\Cavgsca{C_{\rm <sca>}}
\def\Cavgext{C_{\rm <ext>}}
\def\Qavgabs{Q_{\rm <abs>}}
\def\Qavgsca{Q_{\rm <sca>}}
\def\Qavgext{Q_{\rm <ext>}}

\def\G{G_0}
\def\fvoid{f_{\rm vac}}
\def\meff{m_{\rm eff}}
\def\FIGS{}

\def\Rsv{R_{\rm s/V}}
\def\RPp{R_{P/p}}
\def\lmax{\lambda_{\rm max}}
\def\Rv{R_{\rm V}}
\def\Av{A_{\rm V}}
\def\pv{p_{\rm V}}
\def\ebv{E(B-V)}
\def\tauv{\tau_{\rm V}}
\def\Isub{I_{353}}
\def\tausub{\tau_{353}}
\def\Tobs{T_{\rm obs}}
\def\betaobs{\beta_{\rm obs}}
\def\nuzero{\nu_0}
\def\stefan{\sigma_{\rm S}}
\def\Rad{\mathcal{R}}
\def\ebvrad{E(B-V)_\Rad}
\def\ebvps{E(B-V)_{\rm PS}}
\def\albedo{\mathcal{A}}
\def\Psub{P_{353}}
\def\Vband{$\rm V$}
\def\psAvmax{4.4}
\def\PsImax{20}
\def\pmax{p_{\rm max}}

\def\next{43}
\def\npolext{22}
\def\nsed{12}
\def\npolsed{4}

\def\wext{0.70}
\def\wpolext{0.24}
\def\wsed{0.83}
\def\wpolsed{0.51}

\def\fvac{f_{\rm vac}}


\def\chisqextA{0.24}
\def\chisqsedA{3.4}
\def\chisqpolextA{1.48}
\def\chisqpolsedA{31.8}
\def\rchisqA{10.5}
\def\RvA{3.1}
\def\RsvA{4.2}
\def\RPpA{5.7}
\def\IsAvA{1.27}
\def\PsIA{13.3}
\def\pvsAvA{2.9}

\def\rchisqextB{0.42}
\def\rchisqsedB{15.8}
\def\rchisqpolextB{4.92}
\def\rchisqpolsedB{33.3}
\def\rchisqtotalB{15.5}
\def\RvB{2.9}
\def\RsvB{4.1}
\def\RPpB{4.9}
\def\IsAvB{1.12}
\def\PsIB{12.8}
\def\pvsAvB{2.9}

%
\def\rchisqextC{0.53}
\def\rchisqsedC{1.55}
\def\rchisqpolextC{0.49}
\def\rchisqpolsedC{20.7}
\def\rchisqtotalC{6.9}
\def\RvC{2.7}
\def\RsvC{4.0}
\def\RPpC{5.3}
\def\IsAvC{1.23}
\def\PsIC{12.8}
\def\pvsAvC{2.9}

%
\def\rchisqextD{0.41}
\def\rchisqsedD{0.74}
\def\rchisqpolextD{0.20}
\def\rchisqpolsedD{3.8}
\def\rchisqtotalD{1.52}
\def\RvD{3.0}
\def\RsvD{4.0}
\def\RPpD{5.4}
\def\IsAvD{1.23}
\def\PsID{13.1}
\def\pvsAvD{3.0}

\def\amin{a_{\rm min}}
\def\amax{a_{\rm max}}
\def\at{a_{\rm t}}
\def\ac{a_{\rm c}}

\def\aV{a_{\rm V}}
\def\aM{a_{\rm M}}

\def\axisratio{a_\parallel/a_\perp}
\def\stdmodel{PLAWAA42}

\def\mdust{m_{\rm dust}}

\def\incl{\psi}
\newcommand{\healpix}{{\tt HEALPix}}
\def\nside{N_{\rm side}}
\def\mynside{8}
\def\npix{768}

\def\DUSTEM{\texttt{DustEM}}

\title{Dust models compatible with \Planck\ intensity and polarization data in translucent lines of sight}


   \author{V. Guillet\inst{1,4},
	L. Fanciullo\inst{1},
	L. Verstraete\inst{1}, 
	F. Boulanger\inst{1},
	A. P. Jones\inst{1}, 
	M.-A. Miville-Desch\^enes\inst{1},
        N. Ysard \inst{1},
        F. Levrier \inst{2}
        \and
         M. Alves \inst{3}
          }


   \institute{
   1. Institut d'Astrophysique Spatiale, CNRS, Univ. Paris-Sud, Univ. Paris-Saclay, B\^at. 121, 91405 Orsay Cedex, France \\
   2. LERMA, Observatoire de Paris, PSL Research University, CNRS, Sorbonne Universit\'es, UPMC Univ. Paris 06, \'Ecole normale sup\'erieure F-75005, Paris, France \\
   3. CNRS, IRAP, 9 Av. colonel Roche, BP 44346, 31028 Toulouse Cedex 4, France \\
   4. Laboratoire Univers et particules de Montpellier, Universit\'e de Montpellier, CNRS/IN2P3, CC 72, Place Eug\`ene Bataillon, 34095 Montpellier Cedex 5, France
   }

   \date{}

 
  \abstract
{The dust properties inferred from the analysis of \Planck\ observations in total and polarized emission challenge current dust models.}
{We propose new dust models compatible with polarized and unpolarized data in extinction and emission for translucent lines of sight ($0.5 < \Av <  2.5$).}
{We amended the \DUSTEM\ tool to model polarized extinction and emission. 
We fit the spectral dependence of the mean extinction, polarized extinction, SED and polarized SED with PAHs, astrosilicates and amorphous carbon (a-C). The astrosilicate population is aligned along the magnetic field lines, while the a-C population may be aligned or not. }
{With their current optical properties, oblate astrosilicate grains are not emissive enough to reproduce the emission to extinction polarization ratio $\Psub/\pv$ derived with \Planck\ data. Models using prolate astrosilicate grains with an elongation $a/b=3$ and an inclusion of 20\% of porosity succeed. 
The spectral dependence of the polarized SED is steeper in our models than in the data. Models perform slightly better when a-C grains are aligned. A small (6\%) volume inclusion of a-C in the astrosilicate matrix removes the need for porosity and perfect grain alignment, and improves the fit to the polarized SED. }
{Dust models based on astrosilicates can be reconciled with \Planck\ data by adapting the shape of grains and adding inclusions of porosity or a-C in the astrosilicate matrix.}

\authorrunning{V. Guillet et al.}

\titlerunning{Dust model compatible with \Planck\ intensity and polarization data.}

\maketitle

%
%
\section{Introduction}

Dust emission is an important tracer of the interstellar medium (ISM), Galactic or extra-Galactic. It traces the star formation activity and provides estimates of the gas mass \citep{Hild83}. Dust polarized emission allows to measure the orientation of the magnetic field and estimate its strength \citep{CF53}. Dust emission turns into a nuisance for the study of the primorial physics of the universe \citep[CMB B-Modes,][]{PlanckBICEP,PIRL} or of external galaxies, as no field in our galaxy can be considered as free of dust. The characterization of dust properties and their variations through the ISM is therefore an important question for both galactic physics and cosmology.

Dust properties vary from one line of sight to the other \citep{M12}, tracing the proper history of interstellar matter in its lifecycle \citep{J13,Zhu16}. These variations are well characterized in extinction in the diffuse ISM \citep{FM07}. The \Planck\ mission, providing a full-sky survey of the dust total and polarized emission in the submillimeter, has greatly improved our knowledge of dust optical properties in emission, polarized and unpolarized, and allowed to quantify their variations through the diffuse ISM \citep{P2013RXI,PIRXXII,Lapo2015,Gonzalo}. Lately, the balloon-borne observations by BLASTPol characterized the variations of dust polarization properties in higher density regions like the Vela C cloud \citep{Fi15,Ga15,Santos17,Ashton2017}.

The growing amount of available and upcoming dust observations in emission, extinction and polarization on the same lines of sight offers the possibility of a consistent analysis of the full spectrum of dust observables. Differences in \eg\ beam size and path length probed along the line of sight must be carefully taken into account in the line of sight selection and correlation analysis phases. 
If those effects are properly taken into account, one can provide new constraints for dust models \citep[\eg][]{PIRXXI}. 
At high column densities, radiation transfer effects also appear, an effect which can only be solved with a dedicated 3D model \citep{Y12,R16}. 

Existing dust models \citep[\eg,][]{ZDw04,DL07,DF09,MC11,J13, SVB14} are calibrated on the mean extinction curve and the pre-\Planck\ mean spectral energy distribution (SED) observed at high Galactic latitude. They are consistent with cosmic abundances. Some of them \citep{DF09,SVB14} also comply to the mean spectral dependence of polarization in the optical \citep{SMF75}. These models now need to be updated to the new observational constraints extracted from \Planck\ observations of dust polarized and unpolarized emission. \cite{Gonzalo} demonstrated that the \cite{DL07} dust model is not emissive enough per unit extinction, by a factor $\sim 2$, to explain the observed dust SED per $\Av$ at high Galactic latitude. A similar factor was found independently by \cite{Dancanton2015} when comparing dust extinction and emission in M31. In \cite{PIRXXI} (hereafter \PaperI), we showed that the \cite{DF09} dust model, adapted from \cite{DL07}, does not emit enough polarized emission at 353\GHz, by a factor 2.5, compared to its counterpart in polarization in the optical toward stars on translucent lines of sight ($0 < \Av < 2.5$). These new observational constraints all point to the need for more strongly emissive dust grains, especially for the silicate component that is known to be aligned with the magnetic field and therefore responsible for much, possibly all, of the polarization of dust emission and extinction. 

The \cite{J13} dust model, which was built upon theoritical considerations and laboratory data, includes strongly emissive core-mantle grains that can successfully reproduce the dust SED observed by \Planck\ in the diffuse ISM \citep{Lapo2015}. This model is part of a more global renewal of our approach of grain physics \citep[the THEMIS framework, see][]{THEMIS17}, \vincent{which attempts to provide a consistent set of physically motivated predictions for observed variations in dust emission, extinction and scattering from the diffuse to the dense ISM phases \citep{Ko15,Y15HFI,J16coreshine,Y16coreshine}. Defining and calculating the properties of THEMIS grains for polarization is a difficult and long task, owing to their complex structure and composition. This ongoing work will be published elsewhere.}


There is a need for a simple dust model that fits \Planck\ polarized and unpolarized data. 
In this paper, we amend the \cite{MC11} model for polarization. The shape and composition of dust grains are adapted to fit observational constraints in extinction and emission. Such a model can help to analyze polarization data and disentangle between the various physical effects (magnetic field structure, dust alignment, and dust evolution) that contribute to the observed variations of dust polarization properties \citep{Lapo2017}. 

Our paper is structured as follows: Section \ref{DUSTEM} presents the way polarization is implemented in our dust modeling tool, \DUSTEM. Section \ref{Dustprop} details the calculation of the grain absorption and scattering cross-sections for spheroidal grains, and their averaging over the grain spinning dynamics. Section \ref{Data} presents the observational data to be fitted by our models. In Sect. \ref{Modeling}, we detail our dust populations and how we can adapt the shape and porosity of the grains to fit \Planck\ data.
Section \ref{Results} presents four models compatible with \Planck\ polarized and unpolarized data, with or without carbon grains aligned, with porous or composite silicate grains. Follows a short discussion of our results in Section \ref{Discussion}. We conclude with a summary of our results in Section \ref{Summary}.

%
%
\section{Modeling dust polarization with the \DUSTEM\ tool}\label{DUSTEM}

 \DUSTEM\footnote{The version of \DUSTEM\ with polarization can be downloaded from \url{http://www.ias.u-psud.fr/DUSTEM}.}  \citep{MC11} is a fortran code to physically model dust emission and extinction. Combined with its IDL wrapper\footnote{\url{http://dustemwrap.irap.omp.eu}}, it can also fit extinction curves and SEDs. We have amended the \DUSTEM\ and wrapper codes to also model and fit polarization in extinction and in emission. 

\subsection{Modeling of dust alignment in \DUSTEM}\label{DUSTEMPOL}

The polarization of starlight results from the dichroic extinction of light by interstellar dust grains, which are aspherical and aligned along the magnetic field lines \citep{DG51}. The spectral dependence of polarization in the optical shows that while large ($a \ge 0.1\,\mu$m) grains are preferentially aligned, small grains ($a \ll 0.1\,\mu$m) are not \citep{KM95}. Dust emission is also polarized \citep{St66}, with a direction of polarization that is orthogonal to that in extinction.

Interstellar grains behave like spinning tops, gyrating around their axis of maximal inertia. This spin axis precesses around magnetic field lines due to the magnetic moment of the grains. Two main physical processes have been proposed to explain the alignment of grains in the interstellar medium \citep[for a recent review on dust alignment, see][]{ALV15}: by magnetic dissipation \citep{DG51}, or by radiative torques \citep[RATs,][]{DW96,DW97,LH07}. A recent revision of the RAT paradigm shows that magnetic dissipation and radiative torques are both necessary to explain to level of polarization observed in the diffuse ISM \citep{HL16}. 

A physical alignment model must predict the grain dynamics around magnetic field lines as a function of the grain size and of the physical conditions in the environment. This is the case for the alignment process by magnetic dissipation \citep{DG51,SVB14}, but not yet for the RATs, though this is a work in progress \citep{HL16}.
 
The  \DUSTEM\ tool uses a parametric model for dust alignment. For each grain population, grains of a given size $a$ are divided into two components: the grains of the first component are aligned with the magnetic field, while those of the second component have a random orientation and are therefore not polarizing. The fraction of mass of the aligned component, $f(a)$, is a function of the grain radius through 3 parameters $\plev$, $\athresh$ and $\pstiff$:
\begin{equation}\label{Eq-alig}
f(a) = \frac{1}{2}\,\plev\,\left(1+\tanh{\left(\frac{\ln{(a/\athresh)}}{\pstiff}\right)}\right)\,.
\end{equation}
The fraction $f(a)$ increases with $a$ from 0 to $\plev$, with a transition at $a = \athresh$ characterized by a stiffness $\pstiff$. 
For a stiffness $\pstiff=1$, $f(\athresh/2) = 0.2$ and $f(2\,\athresh) = 0.8$. This expression, which gives satisfactory results for all our purposes, also mimics the size dependence of grain alignment expected from RATs \citep{HL16} or from superparamagnetic Davis-Greenstein alignment \citep{Vosh16}.

\subsection{Polarization observables with \DUSTEM}\label{DUSTEMnotations}

The polarization properties of grains do not only depend on the dust alignment efficiency, but also on the magnetic field orientation and its variations along the line of sight. In this article, we model the highest polarization fraction observed in extinction on translucent lines of sight, which most probably corresponds to a situation where the magnetic field \Bmag\ lays in the plane of the sky and is uniform along the line of sight \citep{M07}. To remain general though, we note \Bpos\ the projection of \Bmag\ onto the plane of the sky.

For each grain population,  \DUSTEM\ uses tabulated values in size $a$ and wavelength $\lambda$ of the grain absorption, $\Cabs$, scattering, $\Csca$ and extinction, $\Cext=\Cabs+\Csca$, cross-sections. Different tables are needed. For the unaligned component,  \DUSTEM\ uses the cross-sections $\Cavgabs$, $\Cavgsca$ and $\Cavgext$ calculated for grains in \emph{random} orientation. For the aligned component, cross-sections were calculated for two directions of the $\bold{E}$ component of an incident electromagnetic wave: 
\begin{enumerate}
\item $C_{\rm 1, abs}$, $C_{\rm 1, sca}$ and $C_{\rm 1, ext}$: $\bold{E}$ along \Bpos\,, 
\item $C_{\rm 2, abs}$, $C_{\rm 2, sca}$ and $C_{\rm 2, ext}$: $\bold{E}$ perpendicular to \Bpos\,.
\end{enumerate}
The detailed calculation of those cross-sections is described in Sect. \ref{Dustprop}. For simplicity, we assumed in this article that the grains of the aligned component are in \emph{perfect spinning alignment} (\ie\ that the grain spin axis remains fixed along \Bmag), though our detailed calculations of the grain cross-sections would allow for any grain dynamics. 

The dust total polarization cross-section is the sum of all the contributions of grain size $a_i$ over all grain populations $j$. This yields
\begin{equation}
\sigma_{\rm pol}(\lambda) = \sum_{j,i} n_j(a_i)\,f_j(a_i)\,\left(\frac{\Cextperp-\Cextpara}{2}\right)(j,i)\,.
\end{equation}
The total polarized intensity $P_\nu(\lambda)$ is:
\begin{equation}
P_\nu(\lambda) = \sum_{j,i} n_j(a_i)\,f_j(a_i)\,\left(\frac{\Cabsperp-\Cabspara}{2}\right)(j,i)\,B_\nu(\lambda,T_j(a_i))\,,
\end{equation}
where $T_j(a_i)$ is the temperature of a grain of size $a_i$ from population $j$, as calculated by  \DUSTEM, and $B_\nu(\lambda,T)$ is the associated Planck function. 

\subsection{Extinction and emission}
When dust grains are aspherical, their emission and extinction cross-sections depend in principle on the magnetic field orientation and grain alignment efficiency \citep[\eg][]{HG80,DV10}. However, when the spectral dependence in extinction and in emission are mean values averaged out over many lines of sight at all longitudes, as it is the case in this paper (Sect.~\ref{Data}), dust grains have no prefered orientation. We thererefore assume that grains are all in random orientation for the fitting of the extinction and emission curves:
\begin{eqnarray}
\sigma_{\rm ext}(\lambda) & = &  \sum_{j,i} n_j(a_i)\,\Cavgext(j,i)\,, \\
I_\nu(\lambda) & = &  \sum_{j,i} n_j(a_i)\, \Cavgabs(j,i)\,B_\nu(\lambda,T_j(a_i))\,.
\end{eqnarray}


%
%
\section{Calculation of the cross-sections of spinning spheroidal grains}\label{Dustprop}

Unlike polarization by scattering, which can be produced by non-aligned spherical grains, grains must be both aspherical and at least partially aligned to produce polarization in extinction and in emission. 

\subsection{Grain shape}

From the modeling point of view, the most simple non-spherical shape is the spheroid, which is obtained by rotating an ellipse around one of its axes. If the axis of revolution of the grain is its minor axis, the grain is said to be \textit{oblate}; if the axis of revolution is its major axis, the grain is said to be \textit{prolate}.

We note $\symaxis$ the axis of revolution of the grain, of length $2a$. The two other axes have a length $2b$. 
We define the grain axis ratio $\epsilon = b/a$: $\epsilon > 1$ for oblate grains and $\epsilon < 1$ for prolate grains. 
The radius of the sphere that has the same volume as this spheroid is $\aV = \left(b^2 a \right)^{1/3} = a\,\epsilon^{2/3} = b\epsilon^{-1/3}$. 

To help comparing the properties of oblate and prolate grains, an oblate grain with $\epsilon = 2$ and a prolate grain with $\epsilon = 1/2$ will be said to have the same \emph{elongation}, equal to 2.

\subsection{Grain cross-sections for any inclination angle $\incl$}\label{CrossSection}

In the common view, grains are similar to spinning tops: they spin around their minor axis, which is precessing around magnetic field lines with a nutation angle that can vary with time \citep{DG51}. The grain cross-sections must therefore be calculated for any orientation of the grain with respect to the line of sight.
For symmetry reasons, the grain cross-sections will only depend on the inclination angle, $\incl$, of the grain symmetry axis $\symaxis$ with respect to the line of sight.

Let us consider a spheroidal grain of equivalent radius $\aV$ and axis ratio $\epsilon$, built in a material that has a complex refractive index $m=n+ik$ at a given wavelength $\lambda$. Following the approach of \cite{DV10}, we will calculate the grain cross-sections for 31 values of $\incl$ from 0 to 90\deg\ in steps of 3 degrees.
To that end, we use the T-MATRIX code in its extended precision version \verb?amplq.lp.f?. This is a numerical adaptation of the Extended Boundary Condition Method \citep{Wa71} for non-spherical particles in a fixed orientation \citep{Mi00}.
The code proceeds in two steps: 1) it calculates the $T$ matrix for this particular spheroidal grain characterized by $\aV$, $\epsilon$ and $m$. This is the longest step.
2) Knowing $T$, the code can derive the $2\times2$ complex amplitude matrix $S$ for any orientation of the grain in the laboratory frame (as specified by two Euler angles $\alpha$ and $\beta$), any direction $(\theta_i,\phi_i)$ of the incident beam, and any direction ($\theta_s,\phi_s$) of the scattered beam. This last step, which allows for the calculation of the grain cross-sections, is computationaly fast.

Let us detail this second step. 
We calculate the grain cross-sections for two different orientations of the incident polarization, with either the $\bold{E}$ or $\bold{H}$ component of the incident electromagnetic wave in the plane containing $\symaxis$ and the line of sight \citep{HG80}. 
The extinction cross-sections can be derived from the complex extinction matrix $S$, \ie\ the amplitude matrix when the incident and scattered beams are directed toward the observer \citep[Eqn. (31) and (33) from][]{Mishbook}:
\begin{eqnarray}
\CEext(\incl) & = & 2\lambda \,\Im{(S_{11}(\incl))} \\
\CHext(\incl) & = & 2\lambda \,\Im{(S_{22}(\incl))} 
\end{eqnarray}
where $\Im$ is the imaginary part of the complex. 

The grain scatttering cross-section $\Csca$ is obtained by averaging some components of $S$ over all directions $(\theta_s,\phi_s)$ of the scattered beams \citep{Mishbook}: 
\begin{eqnarray}
\CEsca(\incl) & = & \frac{4\pi}{N}\sum_{s=1}^{N} (\left|S_{11}(\incl,\theta_s,\phi_s)\right|^2+\left|S_{21}(\incl,\theta_s,\phi_s)\right|^2)\\
\CHsca(\incl) & = & \frac{4\pi}{N}\sum_{s=1}^{N} (\left|S_{12}(\incl,\theta_s,\phi_s)\right|^2+\left|S_{22}(\incl,\theta_s,\phi_s)\right|^2)\,.
\end{eqnarray}
The scattering directions are extracted from a \healpix\footnote{For more information, see \url{http://healpix.jpl.nasa.gov} and \url{http://healpix.sourceforge.net}.} map with $\nside=\mynside$ \citep{GHB05}. This provides the zenith and azimuth angles ($\theta_s,\phi_s)$ for $N=\npix$ directions uniformly distributed in space, a sampling good enough for our calculations. 

The absorption cross-sections for $\bold{E}$ and $\bold{H}$ polarizations are given by: 
\begin{equation}
\Cabs^{\rm E,H} = \Cext^{\rm E,H} - \Csca^{\rm E,H}.
\end{equation}

\subsection{Extrapolation in the geometric limit}\label{extra}

The T-MATRIX fortran code \citep{Mi00} fails in the ultraviolet for elongated particles with large $\aV/\lambda$ ratios\footnote{For $\epsilon = 1/3$, this happens when $\aV/\lambda \simeq 3$. For $\epsilon = 1/4$, this happens when $\aV/\lambda \simeq 1.7$.}. The calculation time also increases rapidly with $\aV/\lambda$. To limit the calculation time, we stop our calculations at $\aV/\lambda=4$ for all grains. In case the code fails, we note $\lambda_0$ the shortest wavelength for which the calculation was still successful.
To estimate the absorption and scattering cross-sections at wavelengths shorter than $\lambda_0$, we interpolate between the value calculated at $\lambda = \lambda_0$ and the value predicted by geometric optics.

In geometric optics ($\aV\gg \lambda$), the grain geometrical cross-section, $\Cgeo$, is a function of the inclination angle $\incl$ between the grain's axis of symmetry and the line of sight \citep{B80}:
\begin{equation}
\Cgeo(\incl) = \pi \aV^2 \,\epsilon^{-1/3}\,\sqrt{\epsilon^2\cos^2{\incl}+\sin^2{\incl}}\,.
\end{equation}
To calculate the absorption and scattering cross-section in the geometric limit, we follow \cite{MHK03} and assume that the grain albedo is independent of the grain shape:
\begin{eqnarray}
\Cabs^{\rm geo} & = & 2\,\Cgeo\,\albedo \\
\Csca^{\rm geo} & = & 2\,\Cgeo\,(1-\albedo)
\end{eqnarray}
where $\albedo$ is the grain albedo of a sphere of radius $\aV$, as calculated by the Mie Theory. 

We finally interpolate the absorption and scattering cross-sections between $\lambda_0$ and the \Lyman\ continuum using the linear expression:
\begin{equation}
C^{\rm E,H}_{\rm abs,sca}(\lambda\le\lambda_0) = C^{\rm geo}_{\rm abs,sca} +(C^{\rm E,H}_{\rm abs,sca}({\lambda_0})-C_{\rm abs,sca}^{\rm geo}) \times \frac{\lambda}{\lambda_0}\,.
\end{equation}
Resulting cross-sections are presented in Sect. \ref{Qabs}.

\subsection{Averaging cross-sections over the grain spinning dynamics}\label{spinning}

The grain cross-sections must be averaged over the grain dynamics around magnetic field lines. As this article aims at modeling the highest polarization fraction observed in extinction and in emission, it is reasonable to assume that this corresponds to the case of optimal alignment, \ie\ grains in perfect spinning alignment. Therefore, we ignore the possible nutation and precession of the grain around magnetic field lines, and assume that the spin axis of the grain remains constantly aligned with \Bmag.

The spinning of the grain can be ignored for oblate grains because the spin axis, which is also the axis of symmetry, 
remains parallel to the magnetic field. 
With the notations from Sections \ref{DUSTEMnotations} and \ref{CrossSection}, this yields: 
\begin{eqnarray}\label{Eq-C1C2-POS-obl}
C_1^{\rm oblate} & = & \CE\left(\incl=\pi/2\right) \\
C_2^{\rm oblate} & = & \CH\left(\incl=\pi/2\right)
\end{eqnarray}
This is not the case for prolate grains for which the cross-sections must be integrated over the grain spinning dynamics around its minor axis.
For prolate grains, the grain symmetry axis is perpendicular to the magnetic field during its spinning. Therefore: 
\begin{eqnarray}\label{Eq-C1C2-POS-prol}
C_1^{\rm prolate} & =  & \frac{2}{\pi}\int_0^{\pi/2} \CH\left(\incl\right)\,\d\incl\,. \\
C_2^{\rm prolate} & = &  \frac{2}{\pi}\int_0^{\pi/2} \CE\left(\incl\right)\,\d\incl\,.
\end{eqnarray}

Appendix \ref{MagInclined} gives the expressions for $C_1$ and $C_2$ when the magnetic field is inclined with respect to the plane of the sky, still with perfect spinning alignement. 

\subsection{Grain absorption and scattering cross-sections for random orientations}\label{Random}

Not all grains are aligned with the magnetic field lines. For these grains, the absorption and scattering cross-sections must be calculated in a random orientation. 
Moreover, grains even perfectly aligned with the magnetic field have no prefered orientation with respect to the photons of the interstellar radiation field (ISRF) that heat them. The grain absorption cross-section used to derive the dust temperature must therefore be calculated with no preferred orientation. 

\vincent{For each direction $k$ of the $N=\npix$ directions provided by the \healpix\ $\nside=\mynside$ map, we calculate the inclination angle $\incl_k$ of the grain's axis of symmetry with the line of sight 
and interpolate the $\bold{E}$ and $\bold{H}$ components of the grain absorption cross-section on our library of 31 values for $\incl$ between 0 and 90\deg\ (see Sect. \ref{CrossSection}).} 

The resulting cross-sections for random orientation are given by:
\begin{equation}
C_{\rm <abs>, <sca>} = \frac{1}{N} \sum_{k=1}^N \frac{C^{\rm E}_{\rm abs, sca}(\incl_k)+C^{\rm H}_{\rm abs, sca}(\incl_k)}{2}\,.
\end{equation}

\subsection{Absorption, scattering, and extinction coefficients $Q$}

From the calculated grain cross-sections we define the grain absorption, scattering, extinction, and polarization coefficients, $\Qabs$, $\Qsca$, $\Qext$, $\Qpol$, respectively, as the ratio of the corresponding cross-section to the cross-section of the volume of the equivalent sphere $\pi \aV^2$. 

%
%

\section{Observational constraints for dust in translucent lines of sight}\label{Data}

The spectral data usually used to constrain dust model parameters do not all relate consistently to a specific interstellar environment. 
On the one hand, the extinction curves of \cite{M90} and \cite{F99}, the spectral dependence of the starlight polarization degree \citep{SMF75}, and the amount of extinction per H, $\Av/\NH$, \citep{Bo78,R09}, have been measured toward stars at low Galactic latitude and moderate interstellar extinction ($\Av \sim 1$). On the other hand, the normalization per H, or per Av, of the dust SED has only be made at high Galactic latitude and low extinction through the correlation with the \hi\ 21-cm emission \citep{B96,PIRXVII,P2013RXI} or with the extinction to QSOs \citep{Gonzalo}. 
For the moment, we lack the spectral information at high Galactic latitude in extinction and in polarization by extinction to present a consistent dust model for this particular interstellar medium. 

In \PaperI, we selected 206 suitable lines of sight to stars, at all Galactic longitudes and low Galactic latitude ($|b| \le 30$\deg), to correlate the \Planck\ polarization measures at 353\GHz\ with starlight polarization in the optical taken from the compilation of polarization catalog of \cite{H00}. The \Vband\ band extinction toward those stars spanned a range 0.5 -- 2.5 magnitudes, with a mean of 1.5 mag. These lines of sight, which are not diffuse, should rather be called \emph{translucent}. They do not always cross a translucent cloud however \citep{R09}. For these lines of sight, we found tight correlations between the starlight polarization degree, $\pv$, and the polarized intensity at 353\GHz, $\Psub$, as well as between the polarization fraction in extinction, $\pv/\tauv$, and that in emission at 353\GHz, $\Psub/\Isub$. \PaperI provides the emission to extinction ratios needed to normalize the SEDs of polarized (resp. unpolarized) intensity in units of polarized (resp. unpolarized) extinction: 
\begin{enumerate}
\item $\RPp = \Psub/\pv = [5.4 \pm 0.5]\,\MJysr$,
\item $\Rsv = (\Psub/\Isub)/(\pv/\tauv)= 4.2\pm 0.5$,
\item and consequently, $\Isub/\Av = [1.2\pm0.1]\,\MJysr$. 
\end{enumerate}

\subsection{Extinction $\tau(\lambda)$}\label{ExtData}

As in \cite{MC11}, we use the extinction curve $\tau(\lambda)$ of \cite{F99} for the ultraviolet (UV) and optical (assuming $\Rv = 3.1$), and that of \cite{M90} for the near-infrared (NIR) down to 4\,$\mu$m. 
The normalization of the total extinction $\tau(\lambda)$ per H, needed to compare the mass of dust used in the model with cosmic abundances, is done using $\NH/\Av = 1.87\times10^{21}$ \cmsq\ \citep{R09}. 

\subsection{Optical and near-infrared polarization degree $p(\lambda)$}\label{PolextData}

The UV-to-NIR polarization curve\footnote{The starlight polarization degree, $p$, a percentage, is not to be confused with the polarization fraction of dust extinction, also often noted as $p$ (here denoted as $p/\tau$).}, $p(\lambda)$, is the traditional \cite{SMF75} curve, reviewed by \cite{DAA06} with $\lmax = 0.55~\mu$m and $K = 0.92$
\begin{equation}
p(\lambda)=\pmax\,\exp\left({-K\log^2{\left(\lambda/\lmax\right)}}\right)\,,
\end{equation}
followed by a power-law extension from $\lambda = 1.39~\mu$m up to 4\,$\mu$m with an index of $-1.7$ \citep{MW92,DF09}. 
The value of $p(\lambda)$ at $\lambda=\lmax$, $\pmax$, is set using the maximal observed starlight polarization degree in the optical per unit reddening, $\pv/\ebv = 9\,\%$ \citep{SMF75}. This yields  $\pv/\tauv = 3.15\,\%$ or equivalently $\pv/\Av = 2.9\,\%$. 

\subsection{Dust total emission $I_\nu(\lambda)$}\label{SedData}

\begin{table*} 
\begingroup 
\caption{Mean dust SED per unit $I_{857}$ for the lines of sight with $5 < I_{857} < 20$ \MJysr.} 
\label{TableSEDMAMD}
\nointerlineskip
\vskip -3mm
\footnotesize 
\setbox\tablebox=\vbox{ %
\newdimen\digitlsmwidth 
\setbox0=\hbox{\rm 0}
\catcode`*=\active
\def*{\kern\digitwidth}
\newdimen\signwidth
\setbox0=\hbox{+}
\catcode`!=\active
\def!{\kern\signwidth}
\halign{#\hfil\tabskip=01em&#\hfil&#\hfil&#\hfil&#\hfil&#\hfil&#\hfil&#\hfil&#\hfil&#\hfil&#\hfil&#\hfil&#\hfil\cr
\noalign{\doubleline}
Frequency (\GHz) & 25000 & 12000 & 5000 & 3000 & 2140 & 1250 \cr
$\lambda$ ($\mu$m) & 12 &  25 & 60 & 100  & 140 & 240 \cr 
Instrument & IRAS & IRAS & IRAS & IRAS  & DIRBE & DIRBE \cr 
\noalign{\vskip 5pt\hrule\vskip 3pt}
Total emission $I_\nu$ (\MJysr) &  0.04778   & 0.05092 &    0.3224 &     1.150  &     2.423  &    1.925   \cr 
Statistical uncertainy $\sigma_{\rm s}$ (\MJysr) &   0.00031 & 0.00024 & 0.0017 & 0.005 & 0.006  & 0.002 \cr
Gain uncertainty $\sigma_{\rm g}$ & 0.051 &    0.151  &  0.104  &  0.135 &  0.106 &  0.116 \cr   
Total uncertainty $\sigma_I=\sqrt{\sigma_{\rm s}^2+\left(\sigma_{\rm g}\,I\right)^2} (\MJysr)$ & 0.00246 &  0.00769 &   0.0336  &   0.155   &   0.257   &  0.223  \cr   
\noalign{\doubleline}
Frequency (\GHz) & 857 & 545 & 353 & 217 & 143 & 100 \cr
$\lambda$ ($\mu$m) &  350 & 550 & 850 &1382 & 2098 & 3000 & \cr 
Instrument & HFI & HFI & HFI & HFI & HFI & HFI & \cr 
\noalign{\vskip 5pt\hrule\vskip 3pt}
Total emission $I_\nu$ (\MJysr)& 1.000 &    0.3418  &   0.1022  &  0.02193  & 0.004927  & 0.001559 &\cr 
Statistical uncertainy $\sigma_{\rm s}$ (\MJysr)& 0 &  0.0002 &  0.00010 &  0.00002 & 0.000007 & 0.000003 \cr
Gain uncertainty $\sigma_{\rm g}$ & 0.064 &  0.061 &  0.0078 &  0.0016 &  0.0007 &  0.0009 \cr
Total uncertainty $\sigma_I=\sqrt{\sigma_{\rm s}^2+\left(\sigma_{\rm g}\,I\right)^2}$ (\MJysr) & 0.064 & 0.021 & 0.00080 & 0.00004& 0.000008 & 0.000003 \cr   
 \noalign{\vskip 5pt\hrule\vskip 3pt}
}
}
\endPlancktable 
\endgroup
\end{table*}

We derive a new reference SED, $I_\nu(\lambda)$, for the translucent lines of sight studied in \PaperI. We use the High Frequency Instrument (HFI) \Planck\ maps in their PR2 release available from the \Planck\ Legacy Archive \citep{P2015RVIII}, together with IRIS and DIRBE maps \citep{MAMD-IRIS,DIRBEdata}. Deriving a mean SED per H in translucent lines of sight is a difficult task. In this range of column densities, the 21-cm \hi\ emission is not a good proxy for the total gas column density because of the presence of undetected molecular hydrogen. Instead, we use the 857\GHz\ \Planck\ channel, which has the best signal-to-noise (instrumental as well as generated by CIB fluctuations), to obtain a SED normalized at 857\GHz.
We compute the linear regression of each dust emission channel with the \Planck\ intensity at 857\GHz\ using the \texttt{regress} IDL code, for those lines of sight with $5 < I_{857} < 20~\MJysr$, corresponding to the range $0.5 - 2.5$ mag of \Vband\ band extinction studied in \PaperI. CMB and CIB fluctuations, which are uncorrelated with dust emission, are removed by the correlation analysis, and integrated in the derived statistical uncertainty.
The resulting SED, which is normalized at 857\GHz, is presented in tabulated form in Table \ref{TableSEDMAMD}, with its uncertainty $\sigma_I(\lambda)$ extracted from the correlation. 

\begin{figure}
\includegraphics[width=\hhsize]{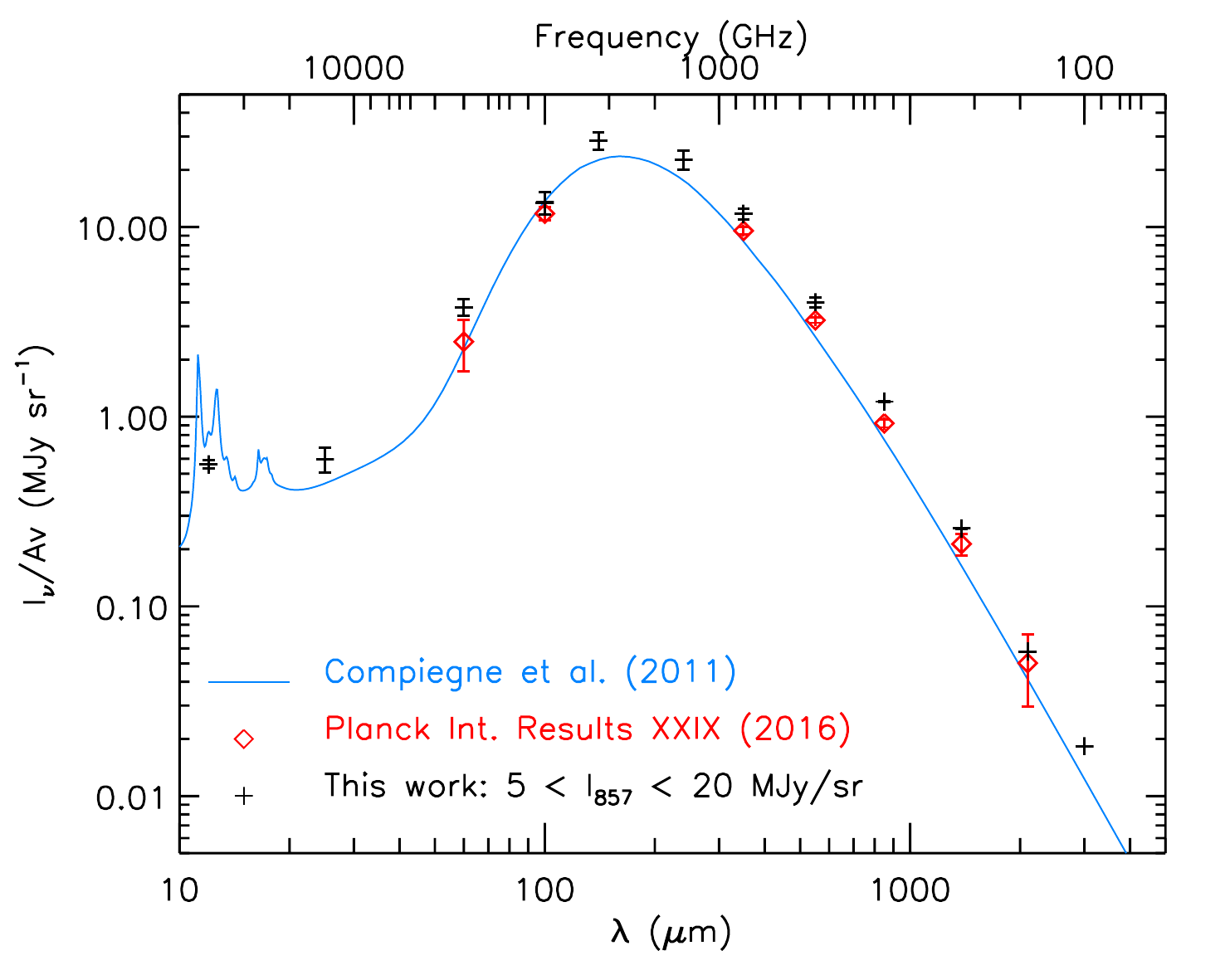}
\caption{(Black) New reference SED for translucent lines of sight ($5 < I_{857} < 20~\MJysr$), converted into a SED per $\Av$ assuming $\Isub/\Av=1.2 \MJysr$. Errors bars include statistical and gain uncertainties, but not the systematic uncertainty of $0.1 \MJysr$ on $\Isub/\Av$ from \PaperI\ (see Section \ref{SedData}). (Red) SED measured in the high Galactic latitude ISM by \cite{Gonzalo}. (Blue) Dust SED from the \cite{MC11} model.}
\label{MAMD}
\end{figure}

The normalized SED is converted into a SED per unit extinction $I_\nu(\lambda)/\Av$ using the ratio $\Isub/\Av~=~ (1.2\pm 0.1)\,\MJysr $ of \PaperI. The uncertainty of $0.1\,\MJysr$, being systematic, is not included in our error bars. The resulting SED is presented in Fig.~\ref{MAMD}, together with the mean SED from \cite{Gonzalo} characteristic of the high Galactic latitude ISM, and the \cite{MC11} model SED. 
All SEDs look very similar in shape, especially in the submillimeter, but the normalization \emph{per unit extinction} emphasizes systematic differences. At high Galactic latitude, $\Isub/\Av=0.9\pm0.1$\MJysr \citep{Gonzalo}, a value which is well accounted for by the \cite{J13} dust model, but not by the \cite{MC11} and \cite{DL07} dust models \citep{Lapo2015}. The present SED, with its higher emission-to-extinction ratio, $\Isub/\Av=1.2$\,\MJysr, characteristical for translucent lines of sight at low Galactic latitude, can neither be reproduced by the \cite{MC11}, nor by the \cite{J13} or \cite{DL07} dust models, even taking into account the systematic error on $\Isub/\Av$ of $0.1\,\MJysr$.

Finally, $I_\nu(\lambda)/\Av$ is converted into an emission per H using $\NH/\Av = 1.87\times10^{21}$ \cmsq\ \citep{R09}. 

\subsection{Polarized SED $P_\nu(\lambda)$}\label{PolsedData}

The wavelength dependence of polarized emission was characterized with \Planck\ HFI data at intermediate, not low, Galactic latitude \citep{PIRXXII}. Variations in the spectral dependence of $I_\nu(\lambda)$ are observed between the diffuse ISM and dense clouds \citep{Y13}, and are therefore expected in $P_\nu(\lambda)$. In the submillimeter, both polarized and unpolarized emission are dominated by the thermal emission of large grains. It is therefore not unreasonable to assume that the ratio of these two quantities, $P/I$, may be less affected by these variations. As a proxy, we therefore assume that the mean wavelength dependence of the dust polarization fraction $P/I$ \citep[Tables 3 and 6 from][]{PIRXXII} is the same at intermediate and low Galactic latitude. 

From the wavelength dependence of $P/I$, we obtain $P_\nu(\lambda)$, the dust polarized emission per H, by multiplying $P/I$ by the $I_\nu(\lambda)$ from Sect. \ref{SedData}. 
From the value of $\Rsv=(\Psub/\Isub)/(\pv/\tauv=4.2$ (Sect. \ref{Data}) and $\pv/\tauv = 3.15\,\%$ (Sect. \ref{PolextData}), we infer a maximal polarization fraction $\Psub/\Isub = 13\,\%$. This is indeed close to the maximal observed polarization fraction in emission at these column densities \citep[][see their Fig.~19 for $\NH\sim 3\times10^{21}$\,\cmsq, corresponding to our mean $\Av\sim1.5$]{PIRXIX}. 

\subsection{Error bars and fitting procedure}\label{fitting}

We use the \DUSTEM\ wrapper based on \verb?MPFIT? \citep{Ma09} to simultaneously fit four curves: the extinction curve $\tau(\lambda)$, the polarization curve in extinction $p(\lambda)$, the total SED $I_\nu(\lambda)$, and the polarized SED in emission, $P_\nu(\lambda)$, which have $n_\tau=\next$, $n_p=\npolext$, $n_I=\nsed$ and $n_P=\npolsed$ data points, respectively, hence a total of $n=81$ data points.

Error bars are set uniformly to 10\,\% of the data for $\tau(\lambda)$ and $p(\lambda)$, except at the \Vband\ band where it is set at 1\,\% because $I_\nu(\lambda)$ and $P_\nu(\lambda)$ are normalized per $\Av$ and $\pv$, respectively. For $I_\nu(\lambda)$, we use the uncertainties derived from our correlation analysis (Section \ref{SedData}), with a lower limit of 1\,\% set to avoid excessive weight from 353 down to 100\GHz. Signal-to-noise ratio for $P_\nu(\lambda)$ is that from \cite{PIRXXII}. 

We choose to give equal weights to each curve $i$ in the fit, \ie\ we decide that the fit is equally constrained by each of the four curves. However, for a given curve, data points with a better signa-to-noise (\eg\ from \Planck) will still have a stronger weight during the fit. To obtain this behaviour, the $\chi^2$ driving the fit is the pondered sum of the four individual $\chi_i^2$ of each curve:
\begin{eqnarray}
\chi^2_i & = & \sum_{j=1}^{n_i}\left(\frac{M_{i,j}-D_{i,j}}{\sigma_{i,j}}\right)^2 \\
\alpha_i & = & \sum_{j=1}^{n_i} \left(\frac{D_{i,j}}{\sigma_{i,j}}\right)^2 \\
\chi^2 & = & \sum_{i=1}^4 \frac{\chi^2_i}{\alpha_i} 
\end{eqnarray}
where $M_{i,j}$, $D_{i,j}$ and $\sigma_{i,j}$ are the model value, the data value and the uncertainty of the data point $j$ from the curve $i$ ($n_i$ data points), respectively.
With this definition, $\chi^2_i/\alpha_i$ is independent of $n_i$. The resulting fit will neither depend on the number of data points, nor on the particular value of the signal-to-noise (10\% here), choosen for $p(\lambda)$ and $\tau(\lambda)$.

%
%
\section{Dust modeling}\label{Modeling}

The submillimeter dust emission probed by \Planck\ HFI, the one we want to model here, is dominated by the emission of large grains ($\sim 0.1~\mu$m), which are also responsible for the polarization of dust extinction and emission. The UV portion of the extinction and polarization curves, together with the mid-infrared emission, are of secondary importance in this article. We therefore decide to simplify the \cite{MC11} description of the grain size distributions in order to limit the number of free parameters and allow for an easier and clearer study of the influence of relevant parameters. 

\subsection{Dust populations}\label{DustPop}

As the PAHs thermal emission is not central to this work, we change the modeling of PAHs in \cite{MC11} from two populations, neutral and positively charged, to one single, neutral, population. 
The size distribution of the two populations of large grains, amorphous carbon (hereafter a-C) and astrosilicate, which were modeled in \cite{MC11} by power-laws with an expononential decay, are now replaced by power-laws. The VSG population is removed and incorporated in the power-law description of the large carbon grain population. Our model therefore entails only three distinct dust populations: 
\begin{enumerate}
\item a population of PAHs, with a log-normal size distribution defined by its adjustable total mass $m_{\rm PAH}$, and its fixed mean radius $a_{\rm PAH}=0.5\,$nm and width $\sigma_{\rm PAH}=0.4$ \citep{DL07};  
\item a population of a-C grains ($\rho = 1.81$ g.cm$^{-3}$), with a power-law size distribution
\item A population of astrosilicate grains ($\rho = 3.5$ g.cm$^{-3}$), with a power-law size distribution.
\end{enumerate}

PAHs do not generally produce a significant polarization feature in the UV, except on two lines of sight \citep{Clay92,Wolff93,Wolff97,HM13}. 
We therefore assume that PAHs are not aligned. 

The optical properties of the a-C population \citep[the BE sample from][obtained through the burning of benzene in air]{Zu96}  were not measured beyond 2 mm, and begin to deviate from a power-law at $\lambda = 1$ mm. As in \cite{MC11}, we extrapolate cross-sections from $800\,\mu$m to 1 cm with a power-law fitted in the range $800-1000\,\mu$m. We also present and discuss in Appendix \ref{discussBE} the strange behaviour of these optical properties when they are applied to small grains in the UV ($\lambda < 0.15\,\mu$m). This has some implications on the size distribution of small carbon grains, but is of minor importance for this article.

The astrosilicates that were used in \cite{MC11} results from the modification by \cite{LD01} (hereafter LD01) of the traditional \cite{DL84} astrosilicates. \cite{LD01} decreased the spectral index of astrosilicates from $\beta = 2$ to $\beta = 1.6$ for 800 $\mu$m $< \lambda < 1$ cm to increase the emissivity of astrosilicate grains. Such an \emph{ad-hoc} modification of the optical properties of astrosilicates is not needed to fit FIR and submm data when one uses amorphous carbon instead of graphite \citep{J13}.
We therefore prefer to use the unmodified, pre-2001, description of astrosilicates \citep[][hereafter WD01]{WD01} with its uniform far-infrared and submillimeter spectral index $\beta = 2.0$. Still, one of our models (model B) will use the LD01 modified astrosilicates for the purpose of comparison.

\subsection{Radiation field intensity $G_0$}\label{G0}


\vincent{The mean radiation field intensity on translucent lines of sight may differ from the one in the diffuse ISM. 
To derive an estimate for the radiation field intensity, namely its scaling factor $\G$, we proposed a method in \cite{Lapo2015} which combines the dust SED with a measure of the dust extinction in the \Vband\ band, that we detail below. 
The dust \emph{radiance}, $\Rad=\int_\nu I_\nu\,\d\nu$, is the total energy (unit W\,m$^{-2}$\,sr$^{-1}$) emitted by dust at thermal equilibrium. $\Rad$ can be calculated by direct numerical integration of the dust SED, or from the observed temperature $\Tobs$, spectral index $\betaobs$, and optical depth at the frequency $\nuzero=353\GHz$ obtained through a fit to the SED \citep{P2013RXI}: 
\begin{equation}\label{Eq-R}
\Rad = \tausub\frac{\stefan}{\pi}\Tobs^4\left(\frac{k\Tobs}{h\nuzero}\right)^{\betaobs}\frac{\Gamma\left(4+\betaobs\right)\zeta\left(4+\betaobs\right)}{\Gamma\left(4\right)\zeta\left(4\right)}
\end{equation}
$\Gamma$ and $\zeta$ are the Gamma and Riemann zeta functions, respectively.

Through the conservation of the energy, $\Rad$ is also equal to the radiative energy \emph{absorbed} by the same dust grains on the line of sight, and is therefore proportional, for a uniform medium in a uniform and anistropic radiation field, to $\G$, to the total amount of dust on the line of sight estimated from the dust extinction $\Av$ in the \Vband\ band\footnote{Like thermal emission in the far-infrared and submilliter, extinction in the \Vband\ band is mainly sensitive to the amount of large ($a \sim 0.1 \mu$m) dust grains on the line of sight.} , and to the dust albedo $\albedo$:
\begin{equation}
\Rad \propto \G \,\Av \,\albedo
\end{equation} 
Following our approach in \cite{Lapo2015}, we use the \emph{radiance per unit extinction}, $\Rad/\Av$, as a proxy for the radiation field intensity.
For the total \Vband\ band extinction on the line of sight, we use the cumulative reddening map, $\ebvps$, based on Pan-STARRS 1 and 2MASS photometry toward a few hundred millions of stars \citep{Green15}. 
For the radiance map, we use improved versions of \Planck\ maps of the dust temperature $\Tobs$, spectral index $\betaobs$, opacity at 353\GHz\ $\tausub$, where the contamination of galactic dust themal emission by cosmic infrared background (CIB) anisotropies has been removed \citep{GNILC}. The radiance map derived from Eq.~(\ref{Eq-R}) is then converted to a reddening, $\ebvrad$, by correlation with the observed interstellar reddening to 200,000 QSOs, as per \cite{P2013RXI}. 

If we make the reasonable assumption that the average radiation field toward QSOs is the ISRF ($\G=1$), we can define the following estimator of $\G$: 
\begin{equation}
\hat{G}_0=\frac{E(B-V)_R}{E(B-V)_{\rm PS}}
\end{equation}
By construction, $\hat{G}_0$ should be close to 1 in the diffuse ISM if the \Planck\ $\ebvrad$ and Pan-STARRS $\ebvps$ maps agree. This comparison was done in \cite{Green15} for the \cite{P2013RXI} radiance map: a good agreement was found betwen $\ebvrad$ and $\ebvps$ below a reddening of 0.3. This  can again be checked for our improved radiance map by controling the value of $\hat{G}_0$ on lines of sight through the diffuse ISM.

\begin{figure}
\includegraphics[width=\hhsize]{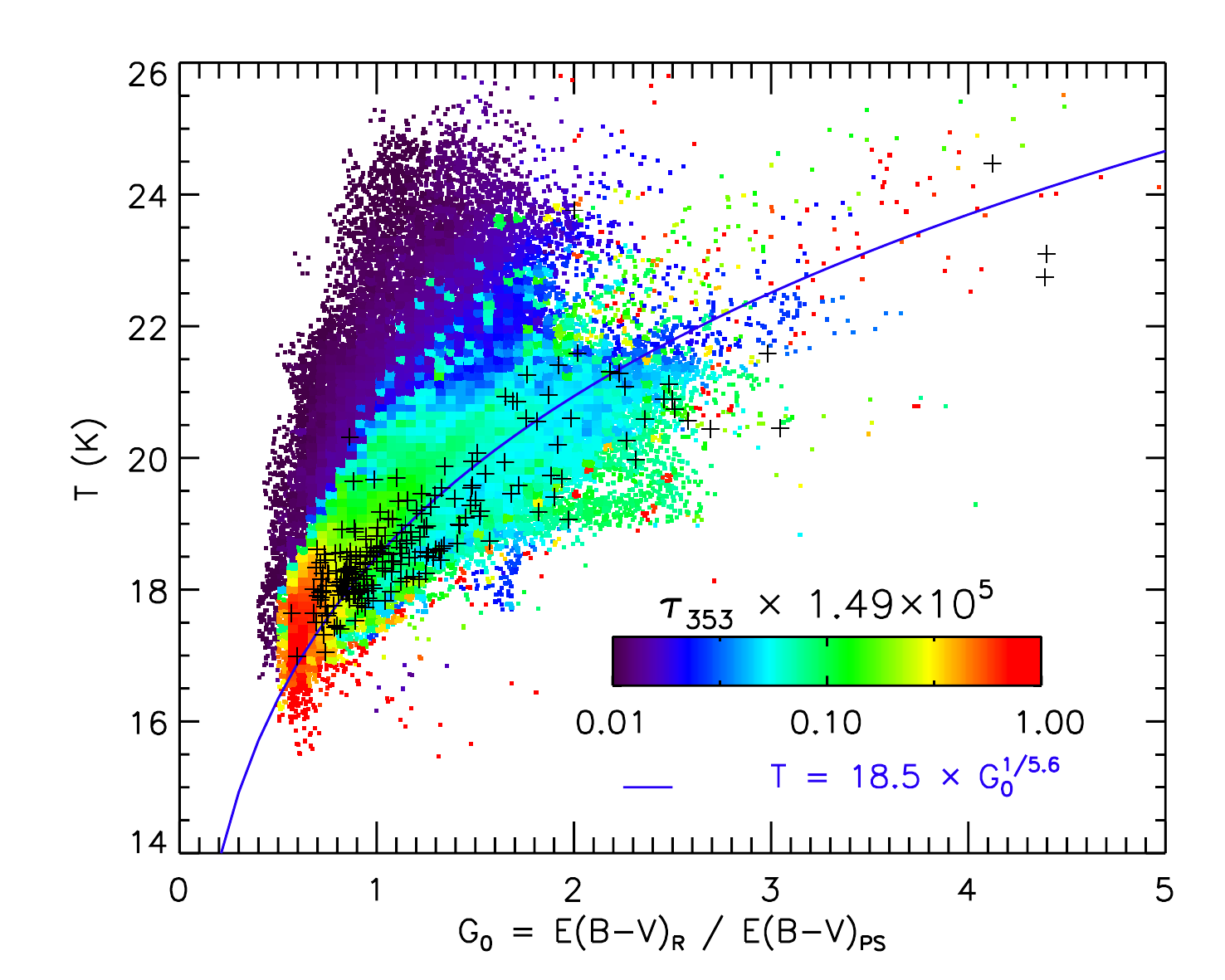}
\caption{\Planck\ dust temperature, $T$, as a function of our estimate for the radiation field intensity, $\hat{G_0}$, colored by the \Planck\ dust optical depth at 353\GHz, $\tausub$, converted to a reddening \citep{,P2013RXI,GNILC}. The blue curve is the prediction for dust with fixed optical properties. Black crosses indicate the position in that plot for the 206 stars from \PaperI.}
\label{T_G0}
\end{figure}

Fig.~\ref{T_G0} presents how the dust temperature $\Tobs$ correlates with our estimator $\hat{G}_0$. Color indicates the mean redening, as derived from the dust optical depth $\tau_{353}$ at 353\GHz. The high latitude diffuse ISM (with an equivalent redening lower than 0.02) has a $\hat{G}_0$ close to 1, as expected. 
Most of the lines of sight to the 206 stars from \PaperI\ also fall around $\hat{G}_0 = 1$. 
Only two lines of sight from our sample sample dense clouds (shown in red, with an equivalent reddening of $\sim 1$). 
The blue line indicates the expected trend $\Tobs=18.5\,\hat{G}_0^{1/5.6}$, assuming that the dust optical properties were uniform in the ISM ($\beta = 1.6$) and fixing $T=18.5\,$K for $\hat{G_0}=1$. Our sample follows this trend line: some lines of sight probe colder dust, some other warmer dust (up to $\hat{G}_0\simeq4-5$ and $\Tobs \simeq 23-24$\,K), but the mean value for the sample is not significantly different from that in the diffuse ISM. Furthermore, we showed in \PaperI\ that the dust properties on those lines of sight with warmer dust did not differ signficantly from the rest of the sample.
}

We therefore decide to fix the radiation field intensity in our models to its reference value, $\G = 1$, as in \cite{MC11}.

\subsection{Calculation of the refractive indices for porous and composite grains}\label{porous}

The introduction of porosity in a grain tends to greatly increase the far-infrared and submillimeter emissivity of grains while only slightly decreasing the optical properties in the optical \citep{J88}. Porosity therefore tends to increase the amount of dust emission per proton or per unit extinction. We will see in Sect. \ref{PolAlone} and \ref{Models} that this is needed. Such introduction of porosity in the grain can be done in a simplified way with the Effective Medium Theory (EMT) that modifies the refractive index of the material by mixing it with that of another material (here the vaccuum). When both components are in comparable volume (we will use porosities of 10 to 30\%), the symmetric Bruggeman mixing rule is recommended \citep{BH83}. Let $m=n+ik$ be the original complex refractive index of the material. We expand the sphere with inclusions of vacuum representing a fraction $\fvoid$ of the \emph{final} volume of the grain, which is then $1/(1-\fvoid)$ times that of the original grain. The mass of the grain is conserved, and so is the radius of the \emph{mass} equivalent sphere, $\aM = \aV\,(1-\fvac)^{1/3}$. Therefore, the material heat capacity is assumed to be unchanged. 

The effective refractive index $m_{\rm eff}$ of the porous grain, from which we can calculate the cross-sections of the porous grain, satisfies the equation \citep{BH83}:
\begin{equation}\label{Eq-Bruggeman}
\fvoid \frac{1-\meff^2}{1+2\meff^2}+(1-\fvoid)\frac{m^2-\meff^2}{m^2+2\meff^2} = 0\,.
\end{equation}
Calculations of grain cross-sections are made for porosities of 10, 20 and 30\,\%.  
To derive the $Q$ coefficients for the porous grains, the calculated cross-sections were normalised by the geometrical cross-section $\pi \aM^2$ of the compact sphere of the same mass. 

In order to better fit the spectral index of \Planck\ polarization data, we will also present a model where the homogeneous astrosilicate population is replaced by a matrix of astrosilicate (refractive index $m_m$) with inclusions of a-C (refractive index $m_i$). The resulting refractive index $\meff$ is obtained using the Maxwell Garnett rule, better suited than the Bruggeman rule when the volume fraction of a-C inclusions, $f_i$, is small \citep{BH83}:
\begin{equation}\label{Eq-Garnett}
\meff^2 = m_m^2\left(1+\frac{3f_i\left(m_i^2-m_m^2\right)}{m_i^2+2m_m^2-f_i\left(m_i^2-m_m^2\right)}\right)\,.
\end{equation}

\subsection{Absorption and scattering coefficients of our dust populations}\label{Qabs}

For each population of astrosilicate and a-C grains, the absorption and scattering coefficients are calculated for grains in perfect spinning alignment with the magnetic field (see Sect. \ref{Dustprop}), for 70 grain sizes equally spaced in log between $a=3$ nm and $a=2$~$\mu$m, for prolate and oblate grains of various elongations (between 1.5 and 4), and for 800 wavelengths between $\lambda=0.09\,\mu$m and $\lambda=1$ cm. 

\begin{figure}
\includegraphics[width=\hhsize]{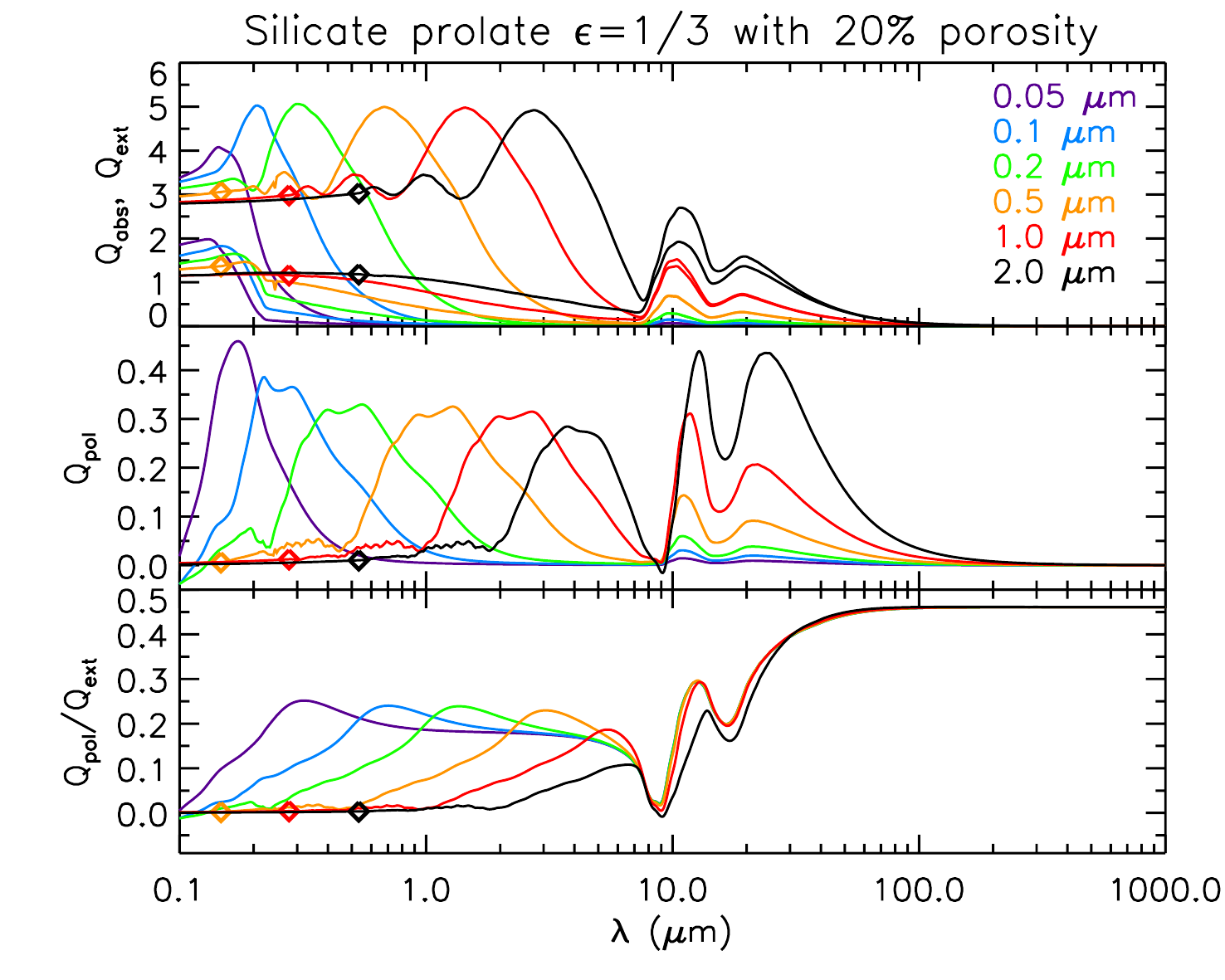}
\caption{Dust absorption ($Q_{\rm abs}$), extinction ($Q_{\rm ext}$), and polarization ($Q_{\rm pol}$) coefficients, as well as the polarization efficiency ($Q_{\rm pol}/Q_{\rm ext}$), as a function of wavelength, for prolate astrosilicates grains with 20\,\% porosity and an axis ratio $\epsilon = 1/3$. The radii $\aV$ of the volume-equivalent spheres are indicated. Black diamonds indicate the wavelength $\lambda_0$ below which coefficients had to be extrapolated (see Sect. \ref{extra}).}
\label{QaSil}
\end{figure}

\begin{figure}
\includegraphics[width=\hhsize]{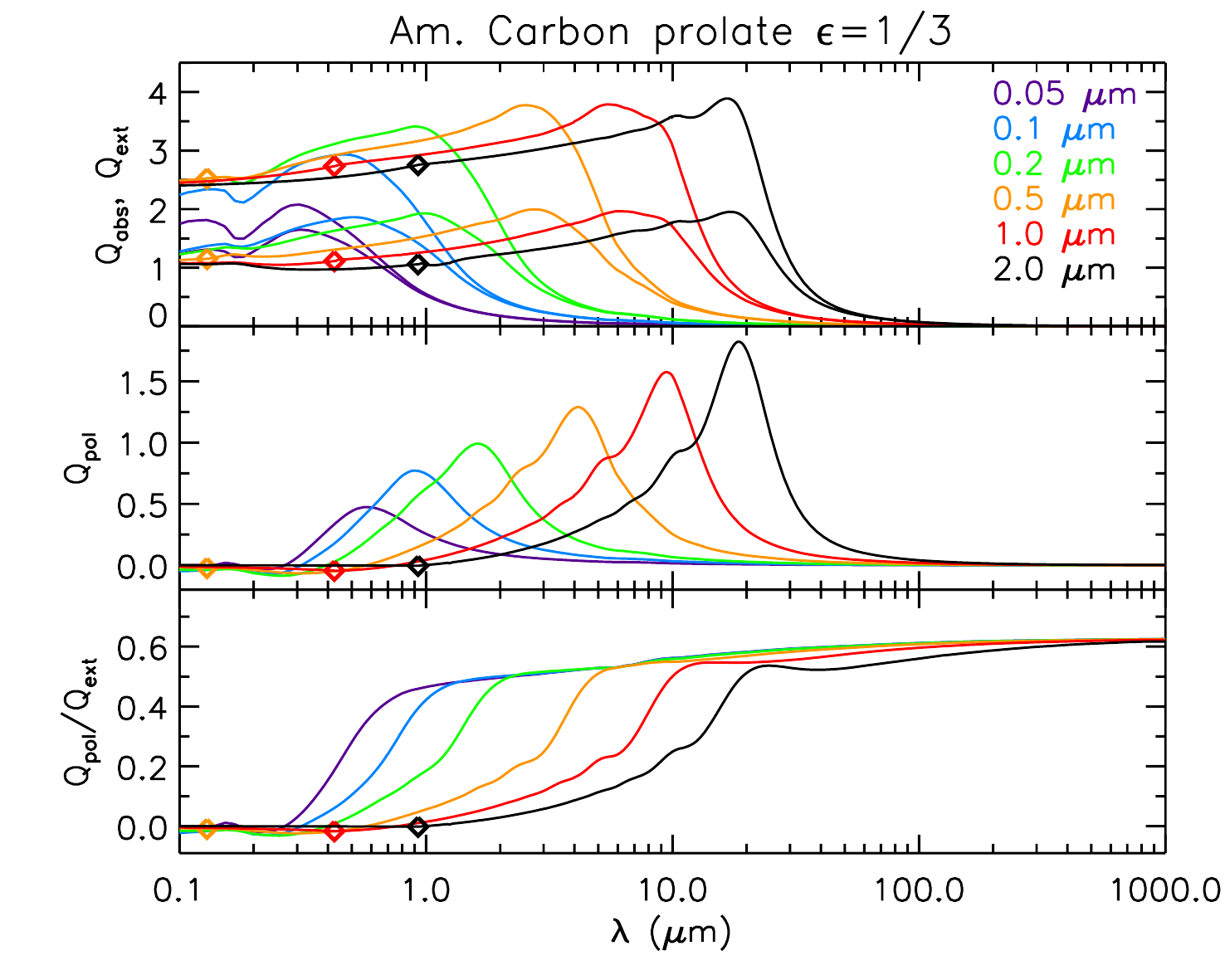}
\caption{Same as Fig.~\ref{QaSil}, for prolate ($\epsilon = 1/3$) a-C grains.} 
\label{QBE}
\end{figure}

Fig.~\ref{QaSil} presents the calculated absorption, extinction and polarization coeffcients of prolate ($\epsilon = 1/3$) astrosilicate grains with inclusions of 20\,\% of porosity, and different sizes from 0.05 to 2 $\mu$m. For the larger grains, the wavelength $\lambda_0$ below which coefficients had to be extrapolated down to the geometric regime are indicated with diamonds (see Sect. \ref{extra}).
Fig.~\ref{QBE} is the same as Fig.~\ref{QaSil} for prolate ($\epsilon = 1/3$) a-C grains.
The impact of our extrapolation down to the \Lyman\ continuum ought to be modest. 
First, the heating of big grains by the ISRF is dominated by the optical and NIR radiation, not by the UV radiation. 
Second, the FUV portion of the polarization curve ($\lambda < 0.15\,\mu$m) is ignored in our analysis. Numerical failures described in Sect. \ref{extra} happen outside of this wavelength range only for grains larger than $0.5\,\mu$m, which are absent from our size distributions. 



\subsection{Constraining the shape and composition of the aligned population}\label{PolAlone}

\begin{figure}
\includegraphics[width=\hsize]{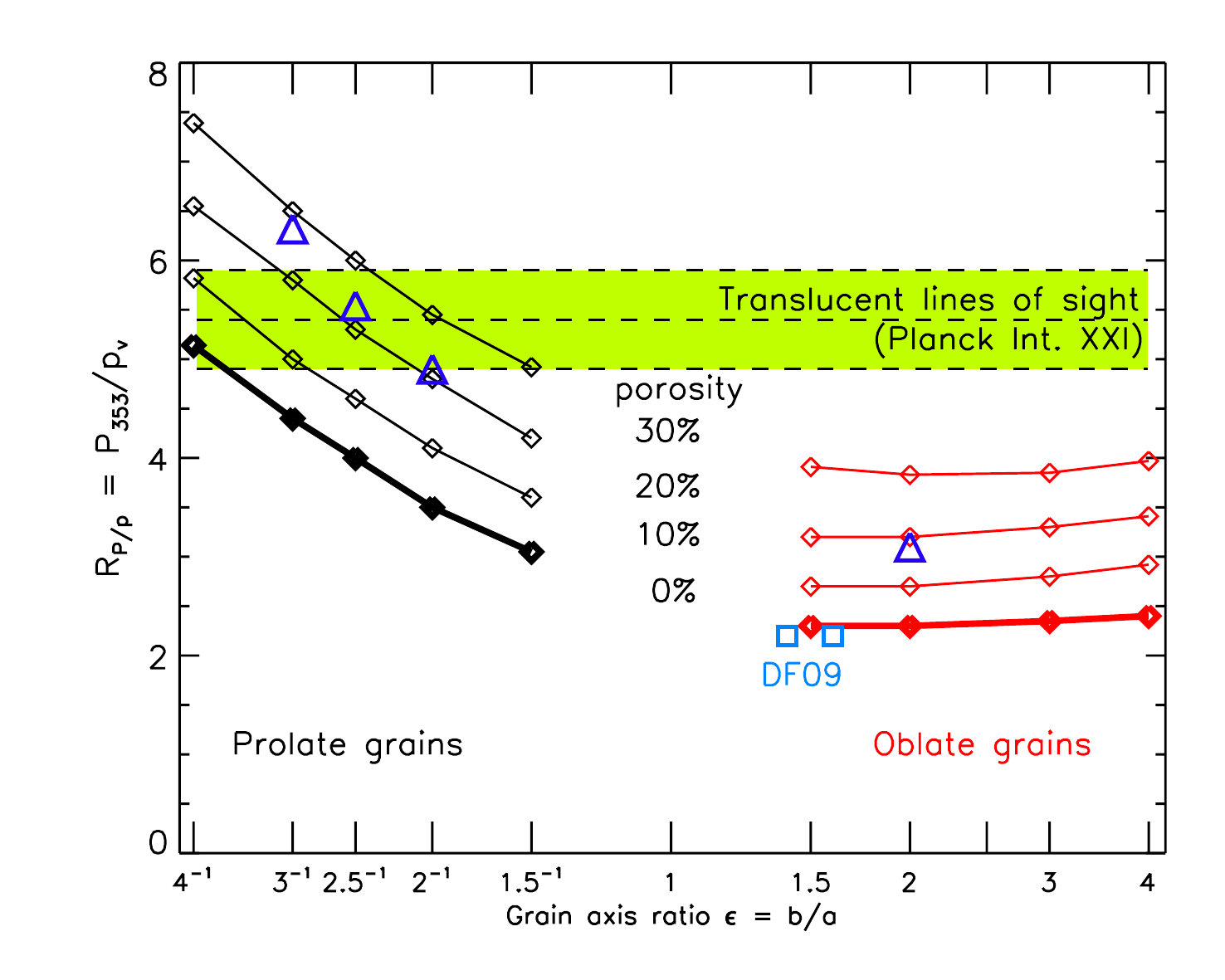}
\caption{Ratio $\RPp$ between the polarized emission at 353\GHz, $P_{353}$, and the polarization degree, $\pv$, measured in extinction in the \Vband\ band, as a function of the grain axis ratio $\epsilon$, for astrosilicates. Labels indicate the grain porosity. The mean value of $\RPp$ in translucent lines of sight (\PaperI), together with its range of uncertainy, is indicated. Blue triangles correspond to an astrosilicate matrix with a-C inclusions (6\,\% in volume). Blue squares correspond to the \cite{DF09} dust models (oblate grains with $\epsilon=1.4$ and $\epsilon=1.6$).} 
\label{RPp_ba}
\end{figure}

\begin{figure}
\includegraphics[width=\hhsize]{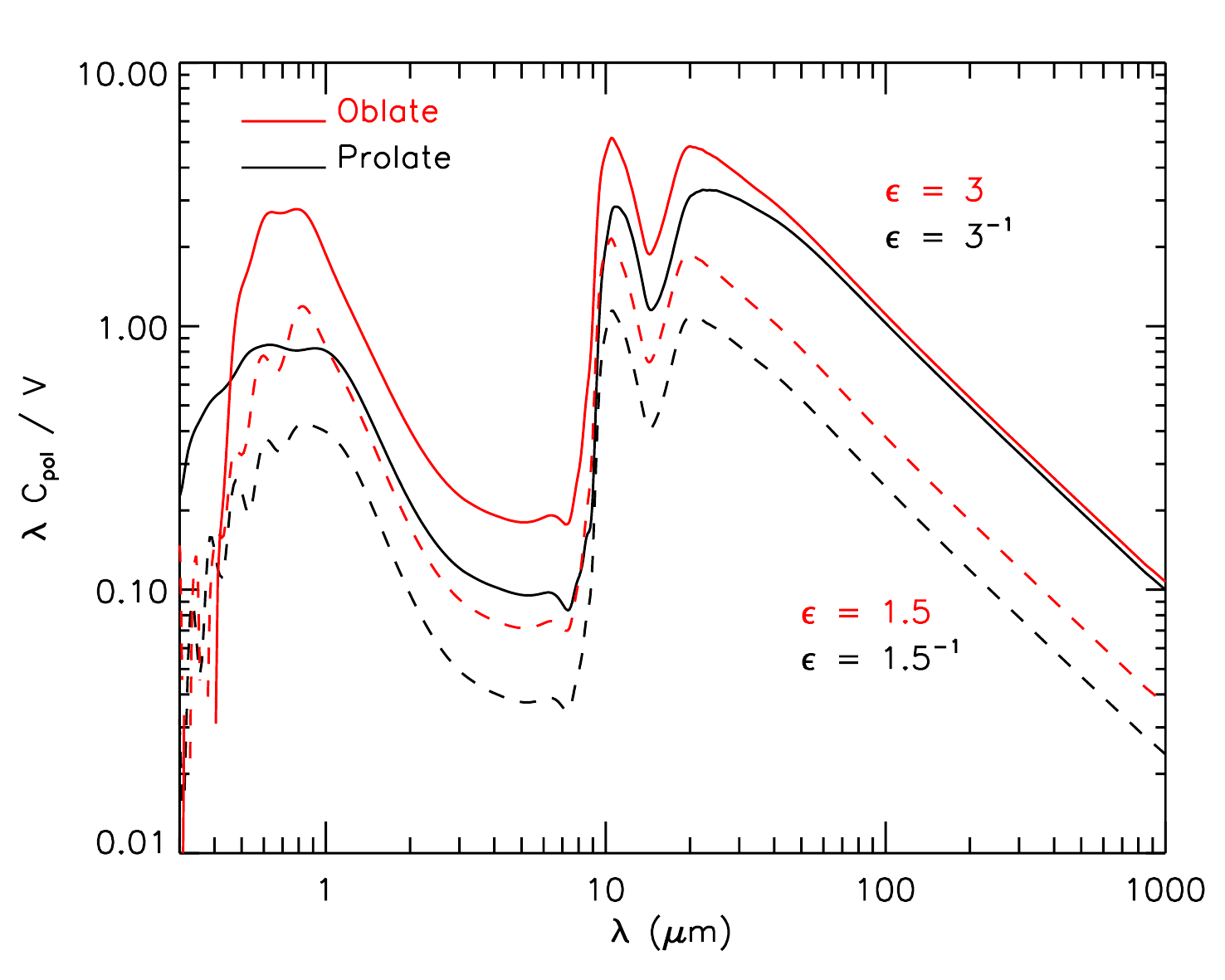}
\caption{Polarization cross-section per unit volume of the corresponding sphere, $\lambda\,\Cpol /V$, as a function of wavelength, for spheroidal grains of equivalent radius $\aV =0.2\,\mu$m. Results are shown in black for prolate grains, and in red for oblate grains, for an elongation of 3 (solid) and 1.5 (dashed).}
\label{whyprolate}
\end{figure}

Polarization measures provide a direct constraint on the aligned grain population alone. In particular, if silicate grains are aligned, and not carbon grains, the polarization curve in extinction provides some insight into the size distribution of aligned silicate grains \citep{KM95}. In this section, we attempt to adapt the shape and the optical properties of the astrosilicate population to reproduce the $\RPp=\Psub/\pv$ polarization ratio. We first test how the grain shape (oblate, prolate) and the axis ratio $\epsilon$ affect $\RPp$. With the  \DUSTEM\ wrapper, we fit the spectral dependence  $p(\lambda)$ alone for prolate and oblate grains of elongation 1.5, 2, 3 and 4. The size distribution of astrosilicate grains is a power-law, with adjustable mass per H $\mdust$, slope $\alpha$ and maximal size $\amax$, and a fixed minimal size $\amin=3$ nm. The alignment parameters, $\athresh$ and $\pstiff$, are also adjustable to fit the spectral shape of the curve, while $\plev$ is fixed to 1. At this stage, no a-C nor PAH populations are added, and we voluntarily ignore other dust observational curves (extinction, total emission, polarized emission). The ratio $\RPp$ is calculated for a mean radiation field intensity ($\G = 1$).

Our results are shown in bold in Fig.~\ref{RPp_ba}. For oblate grains, we find no effect of the grain elongation on $\RPp$: $\RPp \simeq 2.3 \MJysr$ for all elongations up to 4. For prolate grains, we find larger values of $\RPp$, increasing from 3 to 5 \MJysr\ when the elongation increases from 1.5 to 4. The reason for this different behaviour between oblate and prolate shapes is illustrated in Fig.~\ref{whyprolate}. This figure shows the polarization cross-sections of prolate and oblate grains of the same size (0.2 $\mu$m, typical size of aligned grains), for two values of the elongation, 1.5 and 3. For a given elongation, prolate grains polarize less than oblate grains. This difference is more pronounced in the optical than in the submillimeter (hence the higher values of $\RPp$ for prolate than for oblate), and tends to increase with the elongation.
Concerning the effect of porosity, Fig.~\ref{RPp_ba} also shows that it significantly increases the polarization ratio $\RPp$ grains, by a comparable factor for prolate and oblate grains. This effect is almost independent of the grain axis ratio. 

From Fig.~\ref{RPp_ba} we conclude that prolate astrosilicate with elongations betwen 2 and 4, and porosities between 10 and 30\,\%, are able to reproduce the observed $\RPp$ polarization ratio, while oblate grains with elongations from 1.5 to 4, even with porosities as high as 30\,\%, are not. Prolate grains with lower elongation and higher porosities would not allow to achieve the maximal polarization in extinction $\pv/\Av = 3\,\%$. 

We also tested the influence of the introduction of a-C inclusions into the astrosilicate matrix. Carbon inclusions tend to flatten the spectral index of the astrosilicate component, and to increase its emissivity, as needed. Our results for 6\% of a-C inclusions in volume are presented in Fig.~\ref{RPp_ba} as blue triangles. With a-C inclusions, the same value of $\RPp$ can be achieved by the astrosilicate population without the need for porosity, and with a smaller elongation of 2.5. Note that our choice for the volume fraction of a-C inclusions is in agreement with the \cite{J13} dust model, where the amorphous silicate component is covered by an aromatic a-C coating that makes up 6\% of the total silicate mass.

\subsection{Dust models}\label{Models}

It has been demonstrated that the polarization curve in extinction $(p\lambda)$ can be fitted with aligned silicate grains alone, or with both aligned silicate and carbon grains \citep{DF09,SVB14}. We present models for both cases:
\begin{enumerate}
\item Model A: astrosilicate grains are aligned, a-C grains have a random orientation;
\item Model B: like A, but with the optical properties of \cite{LD01} for astrosilicate grains (see Sect.~\ref{DustPop});
\item Model C: both astrosilicate and a-C grains are aligned, with the same alignment parameters $\athresh$, $\pstiff$ and $\plev$ for both populations.
\item Model D: like C, with the homogeneous astrosilicate population replaced by an astrosilicate matrix with 6\% of a-C inclusions, in volume.  
\end{enumerate}

The question whether carbon grains are aligned or not is still debated. From the observational point of view, the absence of polarization of the aliphatic 3.4 $\mu$m band \citep{A99,C06} indicates that aliphatic carbon matter (whether as mantles onto silicate grains, or as an independent population) is not or weakly polarizing. However there are no observational constraints in polarization on \emph{aromatic} carbon grains, the kind that we use here. As demonstrated in \cite{J13}, small aromatic carbon mantles would not produce a spectroscopic signature at $3.3\,\mu$m because of the small carbon mass involved in the mantle compared to the silicate core. We cannot therefore exclude that aromatic carbon grains could be aligned and polarizing, or that silicate grains could be covered by an aromatic carbonaceous mantle like in the \cite{J13} dust model. To be aligned, carbon grains would need some magnetic properties, at least paramagnetic, to precess around magnetic field lines. 

We will assume for simplicity that a-C grains have the same shape and elongation as astrosilicates. As we have seen in Sect. \ref{PolAlone}, prolate astrosilicate grains with an elongation between 2 and 4 are compatible with the observed value for $\RPp$, provided that the astrosilicate matrix is made porous. Another constraint on the elongation of grains, both astrosilicate and a-C, is the observed emission per unit extinction. To achieve the high observed value of the total intensity per unit extinction $\Isub/\Av=1.2\MJysr$ found in \PaperI, the increased emissivity of our porous and elongated astrosilicate population is not enough: we must also increase the emissivity of the a-C population. The elongation of grains, which is necessary to produce polarization when these grains are aligned, can also provide this emissivity increase. 
We find that, for models A, B and C, prolate astrosilicate and a-C grains with an elongation of 3 (and an inclusion of 20\% of porosity for astrosilicates) yield the required emissivity\footnote{The submillimeter absorption cross-section of a-C grains of prolate shape and elongation increase by a factor 3.4 with respect to spheres. WD01 prolate astrosilicate grains of elongation 3 with 20\,\% porosity, by a factor $2.4$. This is in contrast with the optical: their extinction cross-sections are almost unchanged for a-C grains, and decrease only by 20\% for astrosilicates.}, while grains with an elongation of 2 and 2.5 are not emissive enough to achieve the observed $\Isub/\Av$. For model D, an elongation of 2.5 is enough. We further discuss this choice of highly elongated and porous grains in Sect. \ref{Compo}.

%
%
%
\section{Results}\label{Results}

In this section, we compare the best-fit of our four models to the data. 
We recall that the intensity of the radiation field, $\G$, is not\footnote{Fixing $\G$, we give ourselves less flexibility in fitting the shape of the total and polarized SEDs. This is done on purpose, to directly test the optical properties of our dust populations against emission data.} a free parameter: $\G = 1$ (see Sect. \ref{G0}). Each model has 10 free parameters, including 3 for dust alignment. Each fit was obtained assuming that none of the four curves should dominate over the others during the fit (see Sect. \ref{fitting}). Any other assumption would have lead to different solutions.
The $\chi_i^2/n_i$ for each curve $i$, together with the characteristic ratios $\Rv$, $\RPp$, $\Rsv$, $\Isub/\Av$, $\Psub/\Isub$ and $\pv/\tauv$ are given for all models in Table \ref{Tablechi}. All models reproduce the observational emission-to-extinction ratios within the uncertainties (see Sect. \ref{Data}). 

\begin{table*} 
\begingroup 
\caption{Reduced $\chi^2$ for dust models A, B, C and D, together with their characterisc ratios. 
}
\label{Tablechi}
\nointerlineskip
\vskip -3mm
\footnotesize 
\setbox\tablebox=\vbox{ %
\newdimen\digitlsmwidth 
\setbox0=\hbox{\rm 0}
\catcode`*=\active
\def*{\kern\digitwidth}
\newdimen\signwidth
\setbox0=\hbox{+}
\catcode`!=\active
\def!{\kern\signwidth}
\halign{#\hfil\tabskip=01em&#\hfil&#\hfil&#\hfil&#\hfil&#\hfil&#\hfil&#\hfil&#\hfil&#\hfil&#\hfil&#\hfil&#\hfil&\hfil#&\hfil#&\hfil#&\hfil#&\hfil#&\hfil#&\hfil#&\hfil#&\hfil#&\hfil\tabskip=02em#\tabskip=0pt\cr
\noalign{\doubleline}
Model & $\chi^2(\tau(\lambda))/n_\tau$ &$\chi^2(p(\lambda))/n_p$ &$\chi^2(I_\nu)/n_I$ &$\chi^2(P_\nu)/n_P$ &$\Rv$ & $\Rsv$ & $\RPp$ & $\Isub/\Av$ & $\Psub/\Isub$ & $\pv/\Av$ \cr
& & & & & & & (\MJysr) &(\MJysr) & (\%) & (\%) \cr
\noalign{\vskip 5pt\hrule\vskip 3pt}
A & \chisqextA & \chisqpolextA &\chisqsedA&\chisqpolsedA& \RvA & \RsvA & \RPpA& \IsAvA & \PsIA & \pvsAvA \cr
\noalign{\vskip 5pt\hrule\vskip 3pt}
B & \rchisqextB & \rchisqpolextB &\rchisqsedB&\rchisqpolsedB& \RvB & \RsvB & \RPpB& \IsAvB &  \PsIB & \pvsAvB\cr
\noalign{\vskip 5pt\hrule\vskip 3pt}
C & \rchisqextC & \rchisqpolextC &\rchisqsedC&\rchisqpolsedC& \RvC & \RsvC & \RPpC& \IsAvC &  \PsIC & \pvsAvC\cr
\noalign{\vskip 5pt\hrule\vskip 3pt}
D & \rchisqextD & \rchisqpolextD &\rchisqsedD&\rchisqpolsedD& \RvD & \RsvD & \RPpD& \IsAvD &  \PsID & \pvsAvD\cr
\noalign{\vskip 5pt\hrule\vskip 3pt}
}
}
\endPlancktable 
\endgroup
\end{table*}


\subsection{Size distributions and elemental abundances}

\begin{figure}
\includegraphics[width=\hhsize]{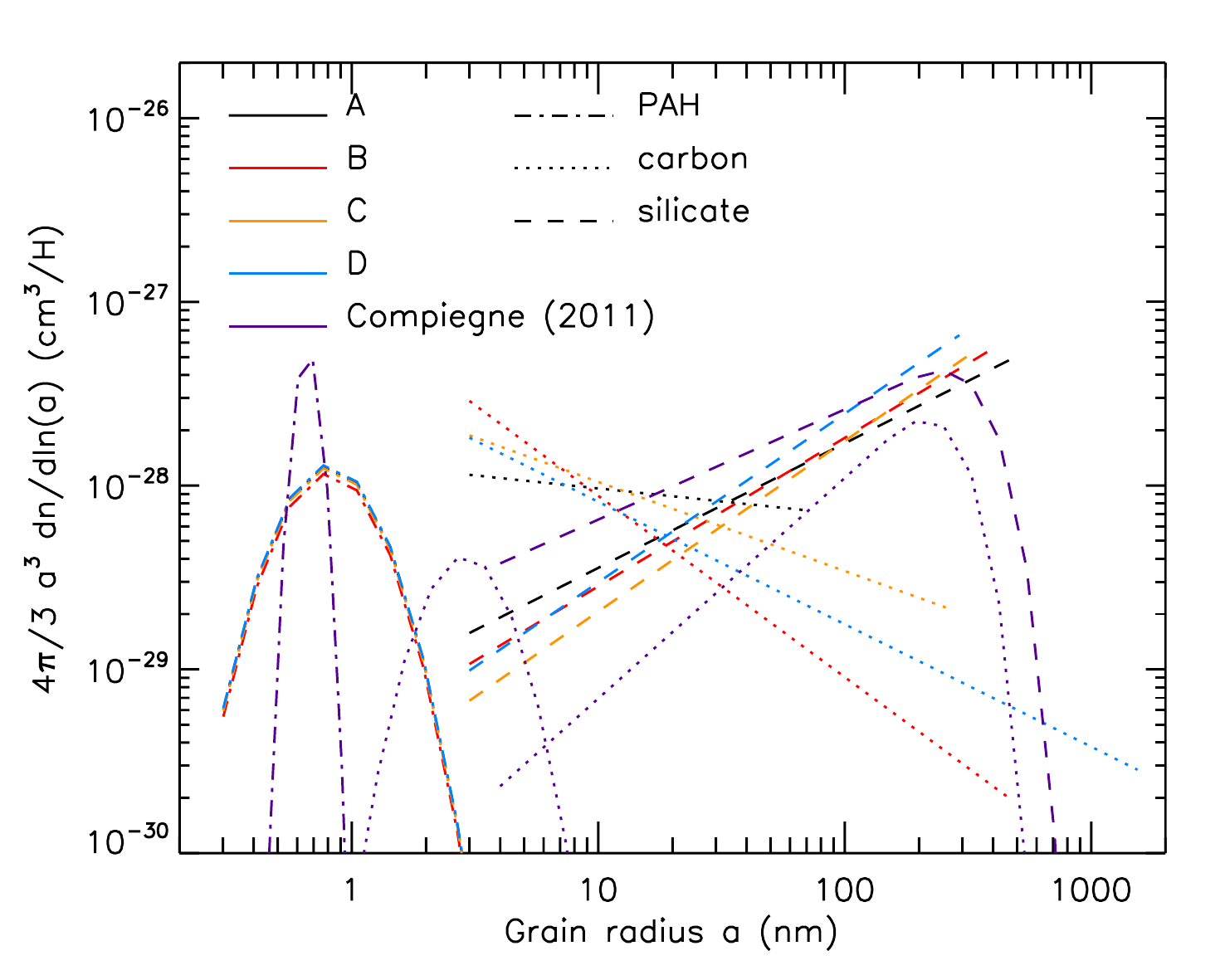}
\caption{Grain size distribution for each population of our four models: PAH (dot-dashed), a-C (dotted) and astrosilicate (dashed). See Table \ref{TableParameters} for the corresponding model parameters. The grain size distribution of the \cite{MC11} dust model is shown for comparison.} 
\label{SDIST}
\end{figure}

Fig~\ref{SDIST} presents the size distribution of PAHs, a-C and astrosilicate grains for our four models, and compare them to the grain size distribution of the \cite{MC11} dust model.
The maximal size for astrosilicate grains ranges from 0.3 to 0.5 $\mu$m depending on the model,
with a slope similar to that of \cite{MC11}. The PAH component is larger, as justified in Section \ref{DustPop}. 
The a-C size distribution is very different from that of \cite{MC11}. It is very steep, with an index close to $-4.5$. This is due to the combination of 3 factors: 1) our models do not include any VSG population \citep{MC11} to reproduce the IRAS mid-IR emission bands; 2) the 60 to 100 $\mu$m color ratio of our SED is higher by a factor $\sim 50\,\%$ than in \cite{MC11} (see Fig.~\ref{MAMD} and Table \ref{TableSEDMAMD}); 3) our a-C population, with its prolate shape (elongation of 3), is more emissive than a-C spheres, and therefore colder. 

\begin{table*} 
\begingroup 
\caption{Size distribution and alignment parameters for our four dust models A, B, C and D described in Sect. \ref{Models}. The radiation field applied to all models is the ISRF ($\G = 1.0$).
}
\label{TableParameters}
\nointerlineskip
\vskip -3mm
\footnotesize 
\setbox\tablebox=\vbox{ %
\newdimen\digitlsmwidth 
\setbox0=\hbox{\rm 0}
\catcode`*=\active
\def*{\kern\digitwidth}
\newdimen\signwidth
\setbox0=\hbox{+}
\catcode`!=\active
\def!{\kern\signwidth}
\halign{#\hfil\tabskip=01em&#\hfil&#\hfil&#\hfil&#\hfil&#\hfil&#\hfil&#\hfil&#\hfil&#\hfil&#\hfil&#\hfil&#\hfil&#\hfil&#\hfil&#\hfil&#\hfil\cr
\noalign{\doubleline}
%
%
Model & Populations & $\epsilon$ & porosity & dust mass & composition &ppm  & $\amax$ & $\alpha$ & $\athresh$ & $\pstiff$ & $\plev$ & \cr 
& & & &  (per H mass) &  & & ($\mu$m) & & ($\mu$m) & & \cr 
\noalign{\vskip 5pt\hrule\vskip 3pt}
%
A &  PAH & \dots & \dots &$7.10\times10^{-4}$ & C &59 & \dots & \dots & \dots & \dots & \dots &   \cr 
& a-C & 1/3 & \dots &$1.32\times10^{-3}$ & C & 110 & $0.070$ &-4.14 & \dots & \dots & \dots  & \cr 
& astrosilicate (WD01) & 1/3 & 20\,\%&$6.52\times10^{-3}$ & MgFeSiO$_4$& 37.9 &  $0.52$ & -3.32 & 0.108 & 0.27 & 1.0 & \cr 
\noalign{\vskip 5pt\hrule\vskip 3pt}
%
%
B & PAH &\dots & \dots&$6.49\times10^{-4}$ & C & 55 & \dots & \dots & \dots & \dots & \dots &   \cr 
&  a-C & 1/3& \dots &$1.32\times10^{-3}$ & C & 110 &   $0.45$ &-4.99 & \dots & \dots & \dots  & \cr 
& astrosilicate (LD01) &1/3 &20\,\% &$6.14\times10^{-3}$ & MgFeSiO$_4$& 35.9 & $0.42$ & -3.19 &  0.110 & 0.34 & 0.86 & \cr 
\noalign{\vskip 5pt\hrule\vskip 3pt}
%
C & PAH & \dots&\dots& $6.98\times10^{-4}$ & C & 58 &  \dots & \dots & \dots & \dots & \dots &   \cr 
&  a-C & 1/3&\dots& $1.56\times10^{-3}$ & C & 130 &  $0.26$ &-4.48 & 0.123 & 0.56 & 1.0 & \cr 
& astrosilicate (WD01) &1/3 &20\,\%& $5.10\times10^{-3}$ & MgFeSiO$_4$& 29.7 & 0.34 & -3.07 & 0.123 & 0.56 & 1.0 & \cr 
\noalign{\vskip 5pt\hrule\vskip 3pt}
%
D & PAH &\dots &\dots& $7.24\times10^{-4}$ & C & 60 &  \dots & \dots & \dots & \dots & \dots &   \cr 
&  a-C &1/3 & \dots& $1.22\times10^{-3}$ & C & 102 &  1.54 &-4.67 & 0.089 & 0.27 & 0.67 & \cr 
& composite astrosilicate/a-C & 1/2.5&0\,\%& $6.20\times10^{-3}$ & MgFeSiO$_4$& 33.9 & 0.29 & -3.08 & 0.089 & 0.27 & 0.67 & \cr
& (6\% of a-C in volume) & &&  & C& 16.0 & \cr
\noalign{\vskip 5pt\hrule\vskip 3pt}
}
}
\endPlancktable 
\endgroup
\end{table*}

The resulting parameters for each model are listed in Table \ref{TableParameters}. 
The elemental abundances consumed by our models are between 30 and 38 ppm for Si, Mg and Fe. These values are of the order of the recommended solar abundances by \cite{Lo03} of 41.7, 40.7 and 34.7, respectively. This is also much less than in the \cite{MC11} dust model which uses 45 ppm of Si, Mg, and Fe. This is primarly a consequence of the porosity introduced in the astrosilicate matrix, and of the prolate elongated shape of a-C and astrosilicate grains.

In translucent lines of sight, depletions are observed to increase with the local average density \citep{Jenkins2009}, so the elemental reservoir available for dust is higher in translucent lines of sight than in the diffuse ISM. This is particularly true for carbon \citep{Parvathi12}. Our models consume between 165 and 188 ppm of carbon, less than the \cite{MC11} model (200 ppm). Models C and D, where a-C grains are aligned, necessitate less silicium, iron and magnesium, but more carbon, than models A and B.

\subsection{Extinction and emission}

\begin{figure}
\includegraphics[width=\hhsize]{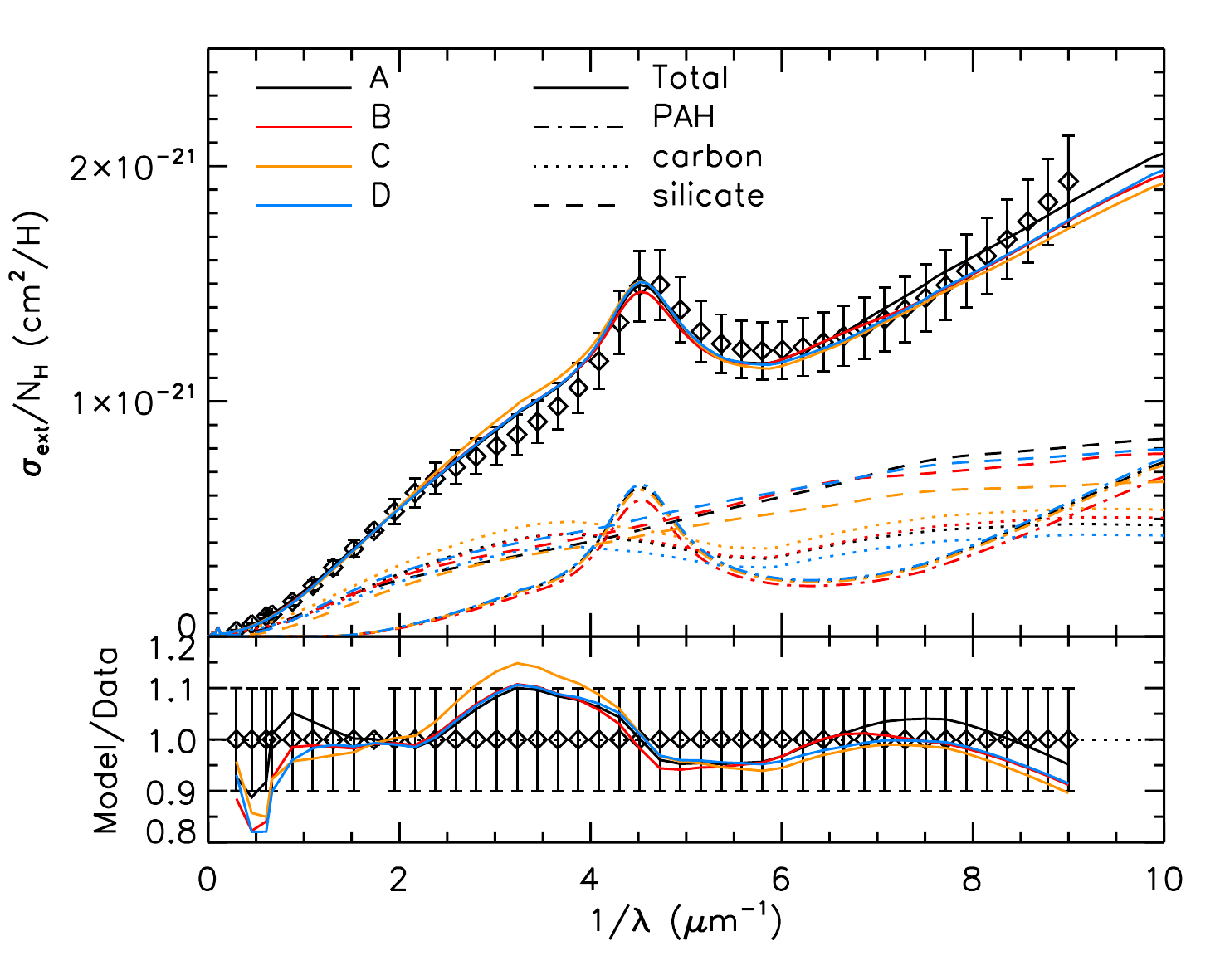}
\caption{Extinction cross-section per H for our four models as a function of inverse wavelength, with the contribution of each component. Error bars are of 10\,\%, except around the \Vband\ band where it is 1\,\%. The lower panel represents a normalized version of the upper panel, where each model has been divided by the data value at each frequency of measure to emphasize the difference between models.}
\label{EXT}
\end{figure}

The extinction curve (Fig.~\ref{EXT}) is reasonably well reproduced ($\chi^2(\tau(\lambda))/n_\tau \ll 1$ with 10\,\% error bars), with an extinction-to-reddening ratio $\Rv$ close to its standart value for the diffuse ISM \citep[$\Rv=3.1$,][]{FM07}, except for model C which uses more a-C and less astrosilicate in mass than models A, B and D.

\begin{figure}
\includegraphics[width=\hhsize]{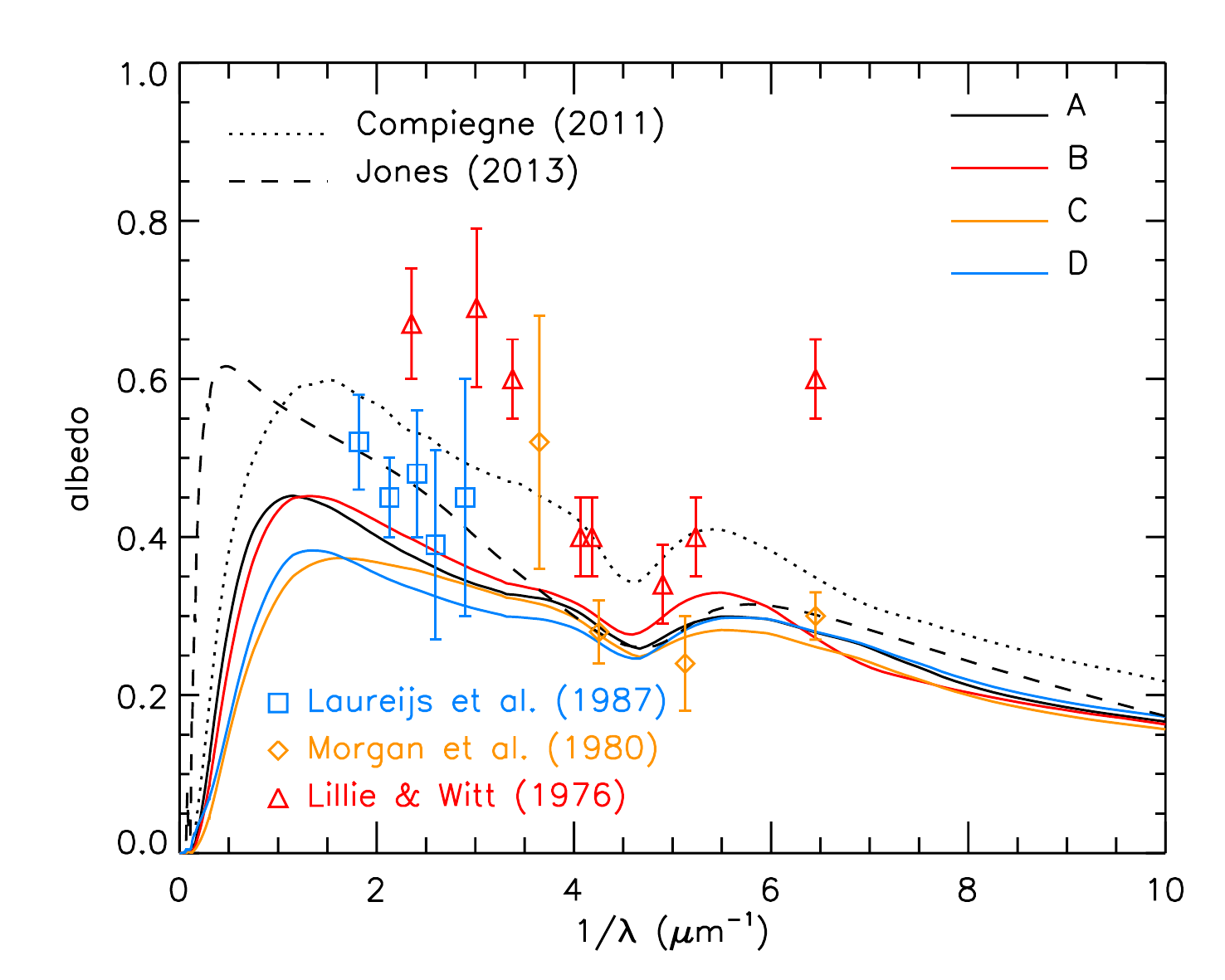}
\caption{Albedo for our four models as a function of inverse wavelength, compared to existing dust models \citep{MC11,J13}. Observational data: red triangles \citep{LW76}, orange diamonds \citep{Morgan80}, blue squares \citep{L87}.}
\label{ALB}
\end{figure}

Fig.~\ref{ALB} shows the albedo for each model and compares it to observational data and predictions from current models. Our models clearly have too low an albedo, in the optical and in the NIR. This is a problem we could not solve with the particular a-C material used here, the BE sample of \cite{Zu96}. See Appendix \ref{discussBE} for a comparison of the extinction properties of the BE sample with other carbonaceous material.

\begin{figure}
\includegraphics[width=\hhsize]{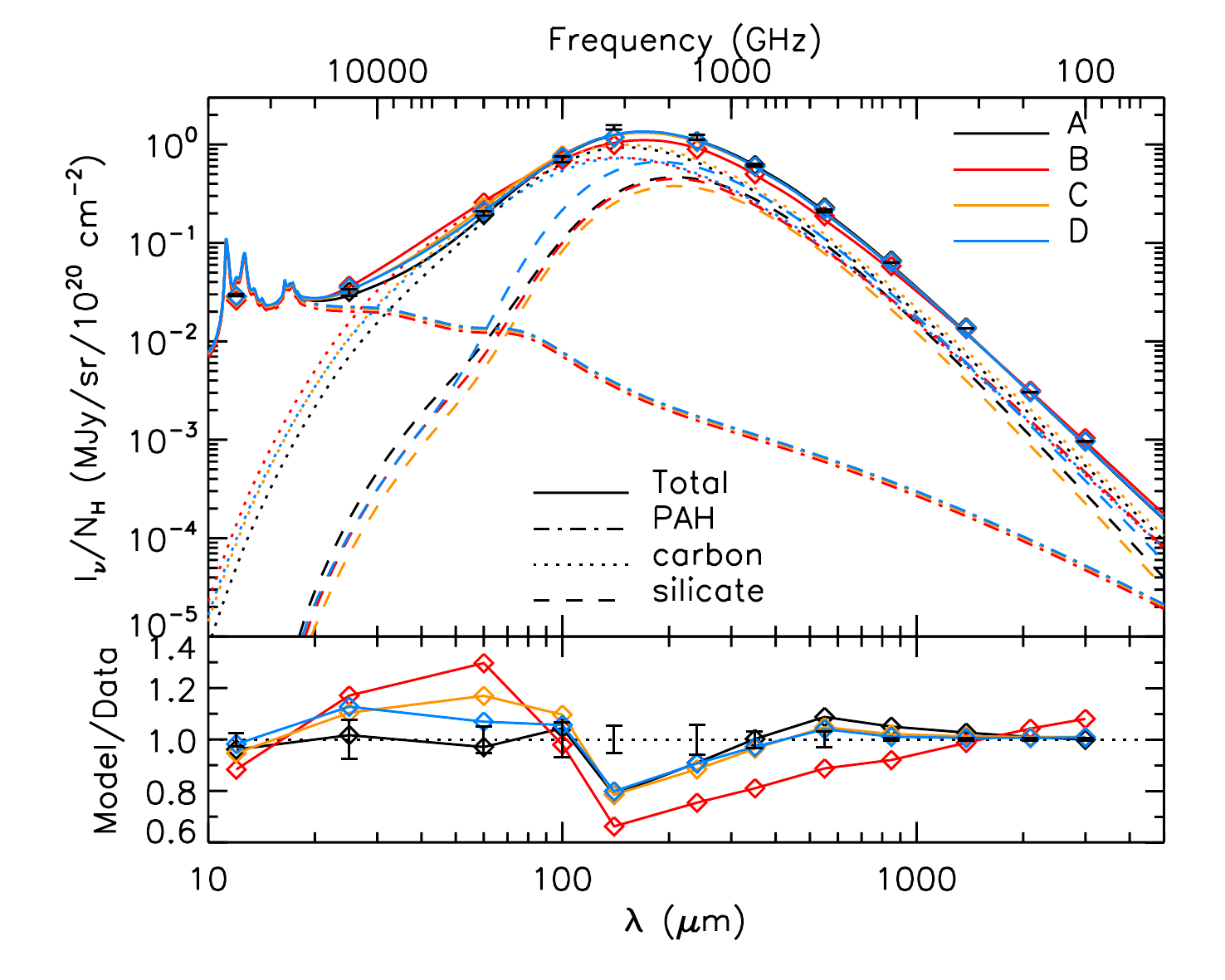}
\caption{Dust emission SED as a function of wavelength, for our four models, with the contribution of each component. Data points and error bars are from Table~\ref{TableSEDMAMD}, with a lower limit of 1\,\% on error bars (see Sect. \ref{SedData}). Diamonds indicate for each model the color-corrected value of the modeled SED in each band of Table \ref{TableSEDMAMD} (approximately $+12\%$ at 353\GHz).}
\label{SED}
\end{figure}

Our models are severly constrained by the dust SED, shown in Fig.~\ref{SED}, especially by its 1\% error bars in the submillimeter (see Table \ref{TableSEDMAMD}). Model B fails to fit the submillimeter emission: the  spectral index of a-C grains ($\beta \sim 1.5$) and of the \cite{LD01} astrosilicates ($\beta \sim 1.6$) are too low to get a good fit.
All models fail to reproduce the dust SED at the DIRBE 140 and 240 micron bands. This was already a limitation of the \cite{MC11} model, that we do not intend to correct here.

\subsection{Polarization}\label{Res-Polar}

\begin{figure}
\includegraphics[width=\hhsize]{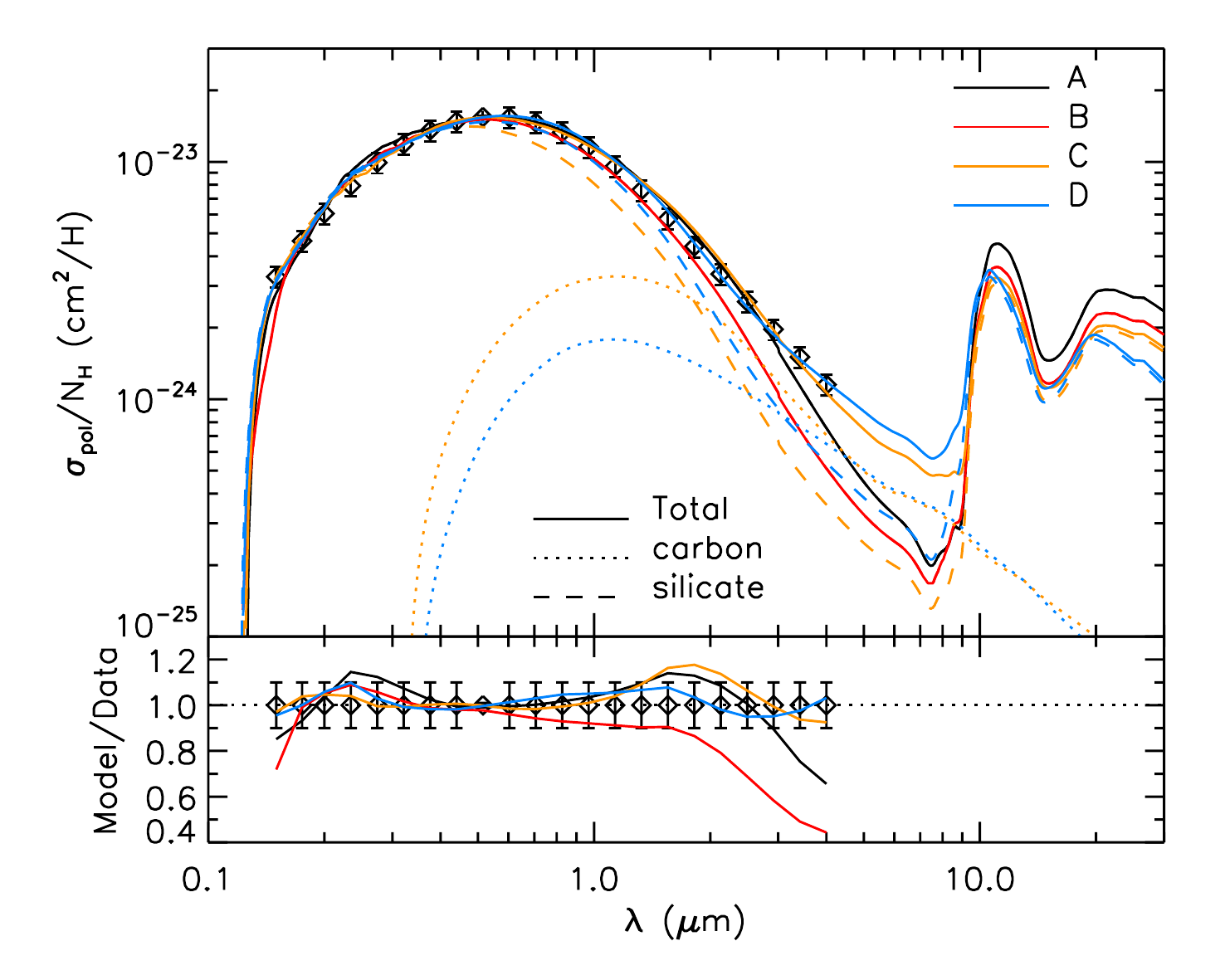}
\caption{Polarization cross-section per H from the UV to the mid-IR, as a function of wavelength, for our four models. Carbon grains produce negative polarization in the UV (\ie\ polarization rotated by 90\deg).}
\label{POLEXT}
\end{figure}

The polarization curve in extinction is shown in Fig.~\ref{POLEXT}. Models A and B, where a-C grains are not aligned, do not reproduce so well the NIR polarization like models C and D, for which a-C grains are aligned. \cite{KM95} had already demonstrated that, by aligning only astrosilicates, the NIR polarization can only be fitted by adding a population of very large ($a\sim 1\,\mu$m) grains. This component, which is not a simple extension of the power-law to large grain sizes, can not be accomodated by our parametrization. In Sect. \ref{LargeGrains}, we discuss the possible contribution of large ($a \sim 1\,\mu$m) astrosilicate grains to the NIR polarization curve by adding an exponential tail to our power-law description of the grain size distribution. 

The peaking wavelength of the $9.7\,\mu$m polarization band is an interesting information to constrain the shape and composition of aligned silicates \citep{LD85,HD95}. We did not intend to fit the profile of this band, but we present their predicted profiles in Fig.~\ref{POLEXT}. We notice that the a-C inclusions in the astrosilicate matrix tends to slightly shift the band to lower wavelengths.

\begin{figure}
\includegraphics[width=\hhsize]{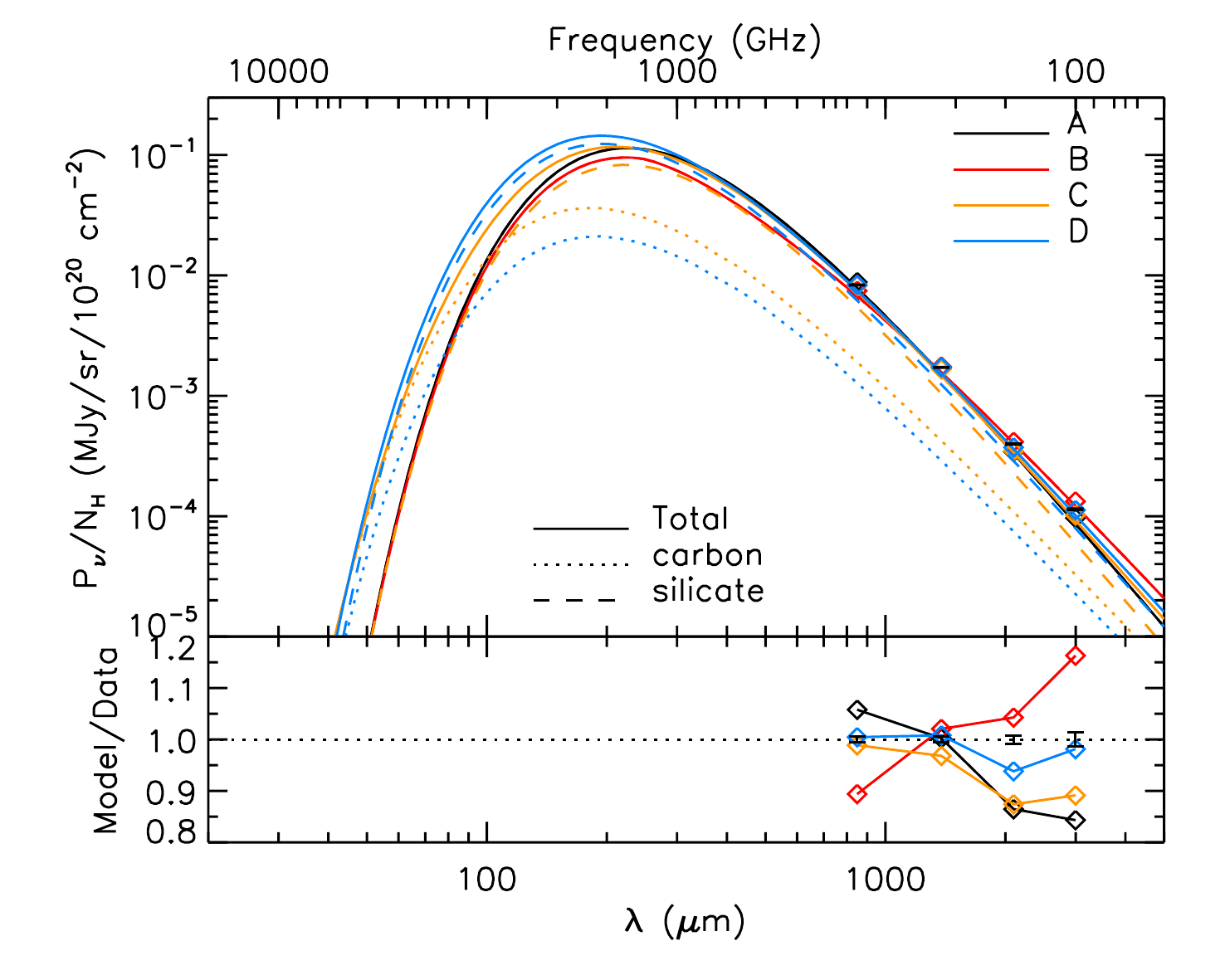}
\caption{Dust polarized SED as a function of wavelength, for our four models, with the contribution of each component. The derivation of data points with their error bars is detailed in Sect. \ref{PolsedData}. Diamonds indicate for each model the color-corrected value of the modeled SED in each \Planck\ HFI polarized band.}
\label{POLSED}
\end{figure}

The polarized SED, shown in Fig.~\ref{POLSED}, is either too steep compared to the data (model A with its spectral index $\beta = 2.0$ for astrosilicates) or too flat (model B with $\beta = 1.6$ for astrosilicates), therefore the bad $\chi^2/n$ values in Table~\ref{Tablechi}.
Aligning a-C grains (Model C) helps to reconcile our model with the low spectral index of the polarized emission observed by \Planck, though only partly.
Model D, with its composite grains mixing astrosilicate with 6\,\% in volume of a-C, performs the best because, in that particular case, all aligned grains have a low $\beta$.

\begin{figure}
\includegraphics[width=\hhsize]{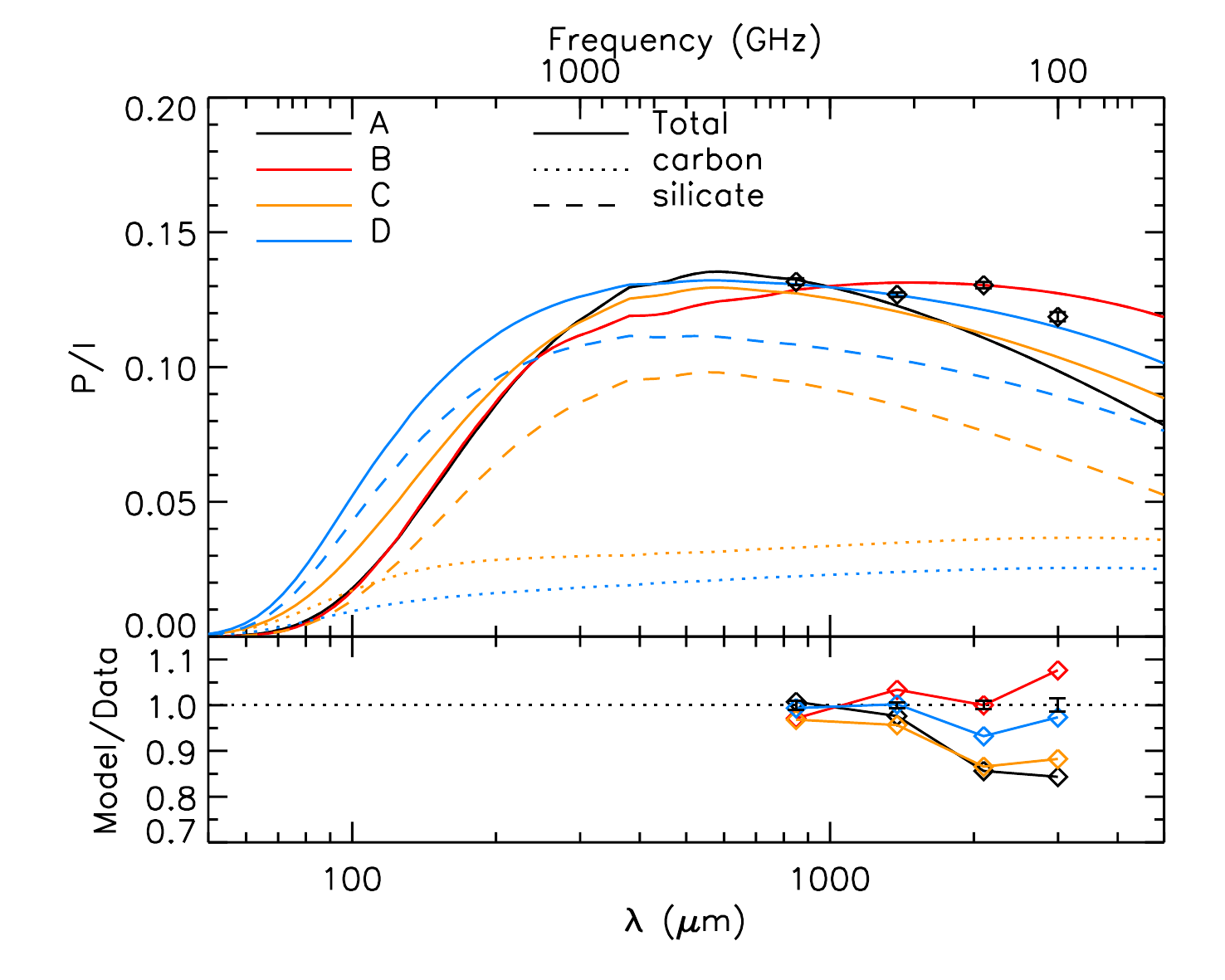}
\caption{Polarization fraction in emission $P/I$ as a function of wavelength, for our four models. Data points and error bars are from \cite{PIRXXII}.}
\label{PsI}
\end{figure}

Fig.~\ref{PsI} presents the polarization fraction in emission. Model B, with a $\beta = 1.6$ spectral index for astrosilicate, yields a reasonable profile, but at the expense of a bad fit to the spectral dependence of the total, $I_\nu(\lambda)$, and polarized, $P_\nu(\lambda)$, intensities (Figs.~\ref{SED} and \ref{POLSED}). Model D is the closest to the data, but still not flat enough. With their aligned a-C grains, models C and D  produce higher polarization fractions in the Wien part of the spectrum (\eg\ at $100\,\mu$m) than model A and B where carbon grains are randomly aligned. This is simply due to the fact that a-C grains and astrosilicate grains with a-C inclusions are warmer than pure astrosilicate grains.

\begin{figure}
\includegraphics[width=\hhsize]{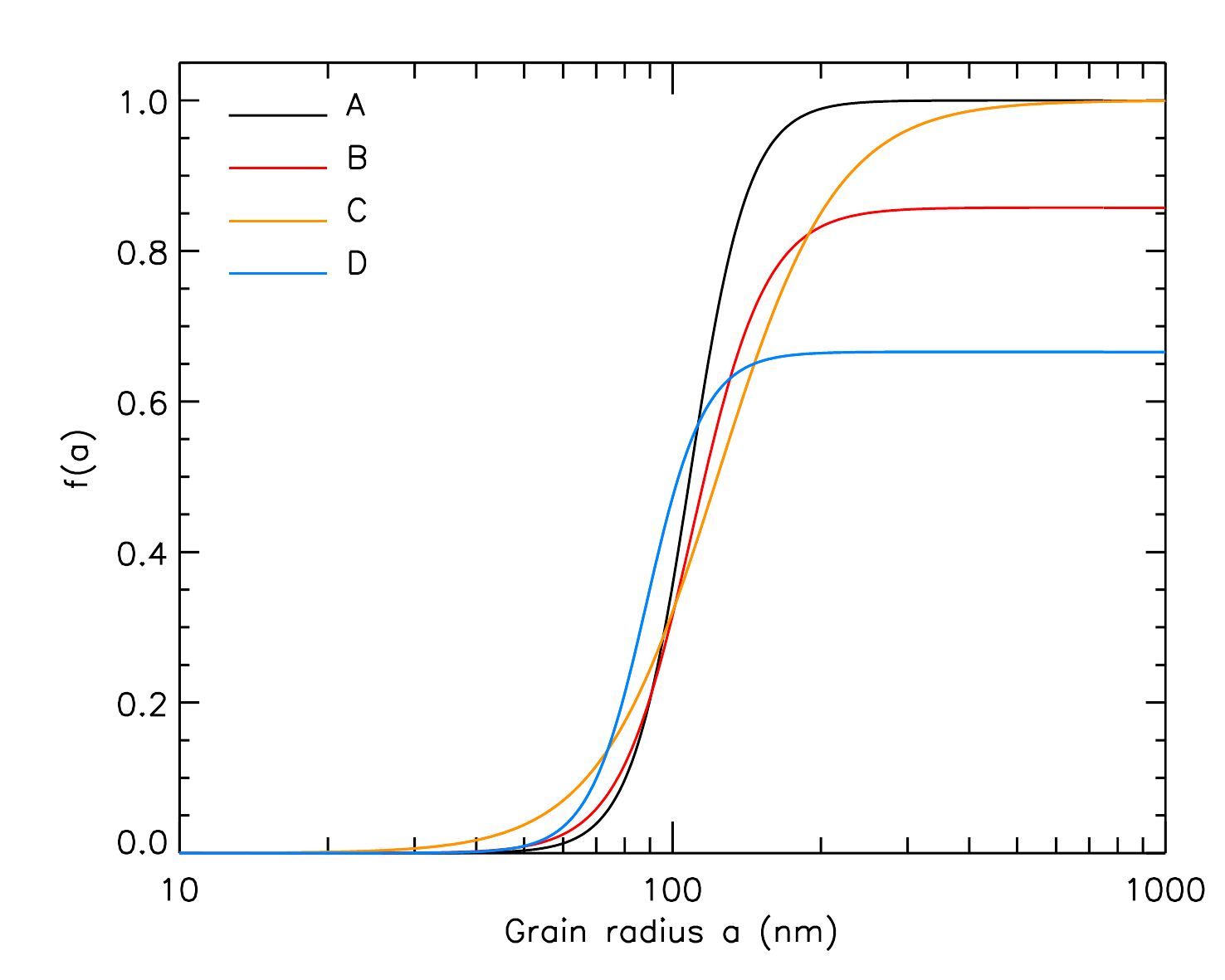}
\caption{Grain alignment efficiency $f(a)$, as a function of grain radius $a$, for our four models. See Eq.~(\ref{Eq-alig}) for the expression of $f$ and Table \ref{TableParameters} for the corresponding alignment parameters.}
\label{FPOL}
\end{figure}

 The corresponding grain alignment functions are presented in Fig.~\ref{FPOL}. The characteristic size for grain alignment ($\athresh$) is close to $0.1\,\mu$m in all models, as expected \citep{KM95,DF09,Vosh16}.
 


%
%
\section{Discussion}\label{Discussion}

In this section, we discuss to which extent our results constrain the size, shape, composition and alignment efficiency of grains.

\subsection{Could large astrosilicate grains also explain the high value of $\RPp=\Psub/\pv$?}\label{LargeGrains}

We did not explore the possible presence of a population of very large astrosilicate grains ($a\sim 1\,\mu$m). Such large and cold grains would increase the polarization ratio $\RPp$ by contributing significantly to the polarized emission at 353\GHz, while producing polarization in the NIR \citep{KM95} but not so much in the optical. 

We model such a population of large astrosilicate grains by an exponential decay attached to the power-law, as per \cite{MC11}, up to $\aV = 2\,\mu$m. This adds three new parameters to the size distribution: $\ac$, $\at$ and $\gamma$ \citep[for a description of those parameters, see][]{MC11}. Following the same approach as in Sect. \ref{PolAlone}, we use \DUSTEM\ to fit the $p(\lambda)$ polarization curve with this single aligned population. We obtain a very good fit ($\chi^2(p(\lambda))/n_p = 0.09$) with a size distribution similar to that of \cite{KM95}. The polarization ratio is however almost unchanged: $\RPp = 5.9\,\MJysr$. We conclude that very large astrosilicate grains ($\aV \sim 1\,\mu$m) do help fitting the NIR polarization curve, but do not contribute significantly to the observed large value of the polarization ratio $\RPp$.

\subsection{What is the maximal grain alignment efficiency in translucent lines of sight?}\label{MaxPol}

The maximal and mean polarization fractions in extinction and in emission are known to decrease with increasing column density \citep[\eg][]{W08,PIRXX,Fi15}. Two interpretations are usually favoured: 1) the grain alignment efficiency, which depends on the environment, may decrease when entering dense clouds due to the extinction of UV photons by dust grains \citep{AP07}; 2) the magnetic disorder along the line of sight may induce depolarization, an effect which increases with the column density \citep{JKD92,PIRXX}. Those effects may combine \citep{JBK15}. We could also mention a third interpretation, supported by observations in unpolarized extinction and emission: 3) the grains composition and shape, and therefore their polarizing efficiencies, could be modified between the diffuse ISM and translucent lines of sight, by accretion and/or coagulation \citep{THEMIS17}. This third option, which greatly increases the complexity of the picture in polarization, is further discussed in Sect. \ref{Compo}. Here, for simplicity, we restrict our analysis to the first two interpretations.

On translucent lines of sight, the column density is high enough for depolarization to occur, and high enough to limit the efficiency of radiative torques. 
If this interpretation is complete, the instrinsic dust polarization fraction, defined as the polarization fraction that would be observed without loss of grain alignment nor depolarization along the line of sight (and therefore, impossible to measure in this medium), must be higher than the one we modeled here, of 3\,\% in extinction and 13\,\% in emission. As a consequence, a dust model for translucent lines of sight should also be able to produce higher polarization fraction than the one observed in that medium, may be up to 20\% in emission - the upper limit observed by \Planck\ at low column densities \citep{PIRXIX}. This condition would exclude models that already necessitate a perfect grain alignement to reproduce the maximal polarization fractions observed in translucent lines of sight, like models A and C, but not the composite model D. In model D, polarization fractions up to $\Psub/\Isub= 20$\,\% in emission and $\pv/\Av=4.5$\,\% in the optical can be achieved\footnote{This linearity between the polarization in extinction and in emission is a consequence of our parametric modeling of dust alignment (see Sect. \ref{DUSTEMPOL}). The limits of this hypothesis will be studied in a dedicated paper.} by increasing $\plev$ from 0.67 up to 1. Such a high polarization fraction in extinction, never observed yet, could only happen at very low column densities ($\Av \ll 0.5$).

\subsection{On the dependence of $P/I$ with the wavelength: models vs observations}

\begin{figure}
\includegraphics[width=\hhsize]{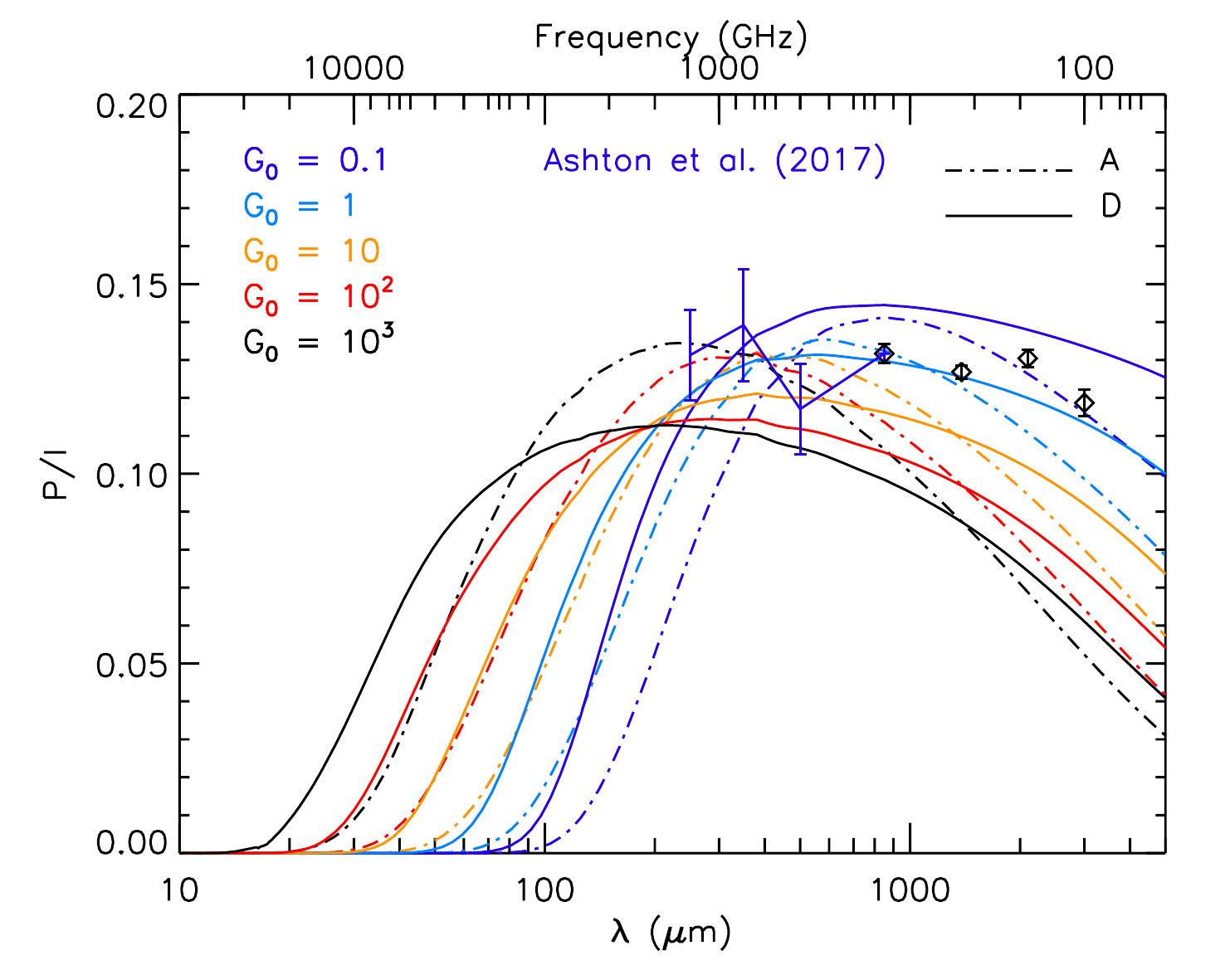}
\caption{Polarization fraction $P/I$ in emission for models D (solid) and A (dashed), for various ISRF intensities $\G$. The spectral shape of the radiation was not changed, just the scaling factor $\G$, as indicated. 
Data points combine \Planck\ \citep[$\lambda \ge 850~\mu$m,][]{PIRXXII} and BLASTPol \citep[$\lambda \le 850~\mu$m,][]{Ashton2017} observations, normalized at 850 $\mu$m.
We cut our spectrum at 5 cm because of the rising contribution to $I_\nu(\lambda)$ of the thermal emission of PAH and other components not modeled here \citep[synchrotron emission, anomalous microwave emission, see][]{PIRXXII}. 
}
\label{SED_G0}
\end{figure}

\vincent{Fig.~\ref{SED_G0} presents our predictions for the wavelength dependence of $P/I$ in models A and D, for a radiation field intensity ranging from $\G = 10^{-1}$ to $\G=10^3$. 
Next to \Planck\ observations, we have added the polarization spectrum recently measured by the BLASTPol balloon experiment in the translucent part ($\Av \sim 2.5$) of the Vela C molecular cloud \citep{Ashton2017}. 
Note that, due to the combined effect of the variation in grain alignement and magnetic field direction, we can  directly compare the spectral dependence of $P/I$ between models and observations, but not its amplitude.
The trend observed by BLASTPol between 250 and 850 $\mu$m is compatible with flatness \citep{Ashton2017}. If we normalize BLASTPol and \Planck\ observations at 850 $\mu$m, we get a flat, possibly slightly decreasing trend over a decade in wavelength, from 250 $\mu$m to 3 mm. With $\G=1$, model D is here again the closest to observations, while model A departs from those data both at short and large wavelengths. 

This flat trend does not seem to be limited to translucent lines of sight. In a previous work \citep{Ga15}, the BLASTPol collaboration already observed a flat spectrum between 250 and 850 microns in the high density regions ($\NH \simeq 2\times10^{22}$\,\cmsq, corresponding to $\Av = 5 - 10$) of the same cloud, heated by the interstellar radiation field ($\G=1$). At even higher column densities, using airborne observations of dust polarized emission in dense regions (among them M17 and M41), \cite{H95} found a maximal polarization fraction at 100 $\mu$m of 9\,\% (for the 99 percentile) for a range $\tau_{100\,\mu{\rm m}} = 0.02-0.2$ of the dust optical depth (or equivalently a range $5-50$ in $\Av$). A polarization fraction of 9\% is obtained with a radiation field $\G \sim 10$ for model D, and with $\G \sim 100$ for model A, which is not unrealistic as those observations were focused on bright regions. 
This peak of $P/I$ at 9\% at 100 $\mu$m is also not so far from what is observed at 353\GHz\ (850 $\mu$m) at the same column densities \citep{PIRXIX}, and does not therefore exclude the possibility of a flatness of the $P/I$ spectrum from the far-infrared to the millimeter.  

In the frame of the astrosilicate-graphite model of \cite{DH13}, the slightly decreasing trend of $P/I$ with the wavelength has been interpreted as an indication of magnetic dipole emission from iron nano-domains embedded in the silicate matrix. In our models, the polarization fraction in emission $P/I$ decreases with the wavelength because the astrosilicate population has a much steeper spectral index ($\beta = 2$) than the a-C population ($\beta \sim 1.5$), that is not compensated by the lower astrosilicate temperature. 

More generally, a dust model with a single (or dominating) homogeneous dust population at thermal equilibrium will have a flat $P/I$ spectrum. In a dust model with 2 populations of large grains, the decrease of $P/I$ with $\lambda$ \emph{in the submillimeter} may indicate that the population with the highest spectral index is more  polarizing (\ie\ better aligned or intrinsically more polarizing) than the one with the lowest spectral index, like is the case in all our models. A flat spectrum would indicate that both populations have the same flat spectrum (\ie\ the same constant $P/I$) and different spectral indices, or different flat spectra (\ie\ different constant $P/I$) but the same spectral index.  
After \cite{Ashton2017}, our modeling also confirms that the flatness of the $P/I$ spectrum \emph{in the far-infrared} probably indicates that the equilibrium temperature of the aligned population does not differ too much from the equilibrium temperature of the unaligned population, unlike models where only silicates are aligned. 

All these observations in translucent lines of sight and dense clouds go againt the natural predictions of most dust models, which were designed for the diffuse ISM and based upon two separate dust populations of large grains with distinct optical properties and alignement efficiencies, like carbon and silicate grains. In the next section, we discuss some hints on the composition of dust grains in translucent lines of sight, and how they will need to be related to the evolution of dust properties from the diffuse to the dense ISM.
}

\subsection{Can we constrain the shape and composition of aligned grains?}\label{Compo}

The shape of interstellar grains is constrained by the spectral dependence of the $9.7\,\mu$m polarization band \citep{LD85}. \cite{HD95} calculated the profile of this band for oblate and prolate astrosilicates and concluded that grains are oblate with an axis ratio $\epsilon < 3$, confirming \cite{LD85} results, in clear contradiction with our models.
However, our results do not demonstrate that grains are necessarily porous, prolate and highly elongated. We did not intend in this paper to determine the shape of interstellar grains, but simply to adapt it to our goal: update the \cite{MC11} dust model to reproduce dust polarized and unpolarized extinction and emission properties on translucent lines of sight, characterized by increased emissivities. This revision of dust opacity is supported by recent laboratory works. \cite{Demyk17} measured the opacity of amorphous silicate from the NIR to the submillimeter, and found it much higher (with a mean factor close to 5) than that of astrosilicates. With such highly emissive material, grains of oblate shape will also be able to reproduce the value of the polarization ratios of \PaperI.

\vincent{Our results also provide new insights on the composition of aligned grains.} Models with aligned a-C grains (models C and D) have the same number of parameters as models A and B with only astrosilicate grains aligned, and perform better in polarization, both in the NIR and in the submillimeter. This is probably indicating that a-C grains may be aligned or that astrosilicate grains may be contaminated by a-C material (in the form of coating or composite aggregates), at least on our translucent lines of sight \citep{J13,KYJ15}.
Model D also gives a possible hint in the same direction: models where aligned grains are composite do not need to invoke perfect spinning alignment ($\plev = 0.67$, see Fig.~\ref{FPOL}), neither very elongated shapes (elongation 2.5), nor porosity to explain observations. 

A physically realistic dust model for polarization should be built on highly emissive, weakly elongated, possibly imperfectly aligned, grains. This approach mixing physically motivated grain shape and composition with optical properties from laboratory data will be pursued in a future work on dust polarization based on the THEMIS approach. 
\vincent{The philosophy of the THEMIS framework \citep{THEMIS17} is not to fit new observations, but to provide a consistent physical framework and a set of laboratory data to relate the observed variations of the dust optical properties to the physical processes affecting interstellar grains during their lifecycle through the various phases of the ISM. 
The THEMIS framework successfully explains the variations of dust emissivity and spectral index from the diffuse to the dense ISM through carbon accretion onto the surface of grains and grain-grain coagulation processes \citep{Ko15,Y15HFI}. It also proposes interesting alternatives to grain growth to explain the phenomenon of coreshine \citep{J16coreshine,Y16coreshine}. Calculating the predictions of the THEMIS model in polarization is not a straightfoward task. First, the calculation of the optical properties of THEMIS core-mantle and aggregate grains takes a few months with DDSCAT \citep{Ko15}, and even longer when one needs to repeat this work for different grain shapes, a key ingredient for polarization. Second, the evolution of the shape and composition of dust grains should be determined in relation to the physical evolutionary processes affecting dust grains through the ISM, particularly from the diffuse to the dense phase, consistently with polarization observations. This is a work in progress that will be published elsewhere.}

%
%
\section{Summary}\label{Summary}

\DUSTEM, a modeling tool for dust extinction and emission \citep{MC11}, was amended to model dust polarization in extinction and in emission. 
The modeling of grain alignment in \DUSTEM\ is parametric: the fraction of aligned grains is a function of the grain size through three parameters. Grains cross-sections were calculated for spheroids of prolate and oblate shape, in perfect spinning alignment when they are aligned, and in random alignment when they are not aligned.

We gathered a consistent set of observational data characterizing the spectral dependence of the mean dust extinction, maximal polarization in extinction, mean SED and maximal polarized SED on translucent lines of sight ($\Av = 0.5-2.5$). Polarized (resp. total) emission curves were normalized per unit polarized (resp. total) extinction in the optical using the submillimeter-to-optical ratios measured in \PaperI: $\Psub/\pv=[5.4\pm0.5]\MJysr$ (resp. $\Isub/\Av = [1.2\pm0.1]\MJysr$).
\vincent{Combining the \Planck\ radiance map \citep{GNILC} and the Pan-STARRS reddening map \citep{Green15}, We estimated the radiation field intensity in our sample, and found it compatible with the insterstellar standard radiation field. We therefore assumed $\G = 1$ in our modeling.}

Our models, derived from the \cite{MC11} dust model, use PAHs, amorphous carbon \citep[BE,][]{Zu96} and astrosilicates \citep{WD01}. In the particular case where only astrosilicates are aligned, we demonstrated that the modeled polarization ratio $\Psub/\pv$ increases with the grain elongation in the case of prolate grains, but remains constant at $\Psub/\pv\simeq 2.3\MJysr$ in the case of oblate grains. The inclusion of porosity in the silicate matrix increases the $\Psub/\pv$ ratio in a similar manner for prolate and oblate grains. We found that porous prolate astrosilicate grains, of elongation $a/b$ between 2 (with a porosity of 30\%) and 4 (with a porosity of 10\%), could reproduce the high polarization ratio $\Psub/\pv$ from \PaperI, while no oblate grains could. This result applies to all dust models using aligned astrosilicates, independently of the particular material choosen for the other dust populations. Conclusions for other aligned materials may differ in the grain shape, elongation and porosity. We also showed that the presence of a population of very large ($a\ge 1\,\mu$m) astrosilicate grains only marginally affects this polarization ratio, while the alignment of a-C grains makes it slightly decrease. 

The observed submillimeter-to-optical ratios $\Psub/\pv$ and $\Isub/\Av$ can be reproduced if the grain shape of a-C and astrosilicate is prolate with an elongation $a/b = 3$, and astrosilicates are made porous with 20\% of porosity. We presented four models compatible with our data set and with cosmic abundance constraints, with a-C grains aligned or not. The spectral dependences of the extinction curve, of the polarization curve in extinction, and of the total SED are well reproduced. 
All models however generally suffer from a too low albedo in the optical. The polarized SED is not well reproduced when only astrosilicates are aligned. Models perform slightly better when a-C grains (with their spectral index $\beta \sim 1.5$) are also aligned. A composite model where the astrosilicate component is mixed with a small fraction of a-C inclusions (6\% in volume) performs even better, while necessitating less elongated ($a/b=2.5$), and less efficiently aligned, grains. These new insights into the properties of dust support the scenario of a dominant dust component of carbon-coated silicates as proposed by \cite{J13}. 

\vincent{
The value and wavelength dependences of $P/I$ in our models are globally consistent with far-infrared and submllimeter observations in translucent and dense regions \citep{H95,Ga15,Ashton2017}, with a better agreement for our composite model with a-C grains aligned.}
Because of the lower spectral index of the amorphous carbon population ($\beta \sim 1.5$), models reproduce the decrease of the polarization fraction in emission $P/I$ with the wavelength, without the need for the magnetic dipole emission of metallic domains in the silicate matrix proposed by \cite{DH13}. 

The \DUSTEM\ code, documentation, and input files for our models are available from \url{http://www.ias.u-psud.fr/DUSTEM}.
With their simple power-law size distributions, and reduced number of free parameters, our models can be used to analyze multiwavelength polarized and unpolarized data, and study the influence of relevant parameters \citep{Lapo2017}.

\appendix

\section{Inconsistency in the use of the BE sample for very small grains}\label{discussBE}

\begin{figure}
\includegraphics[width=\hhsize]{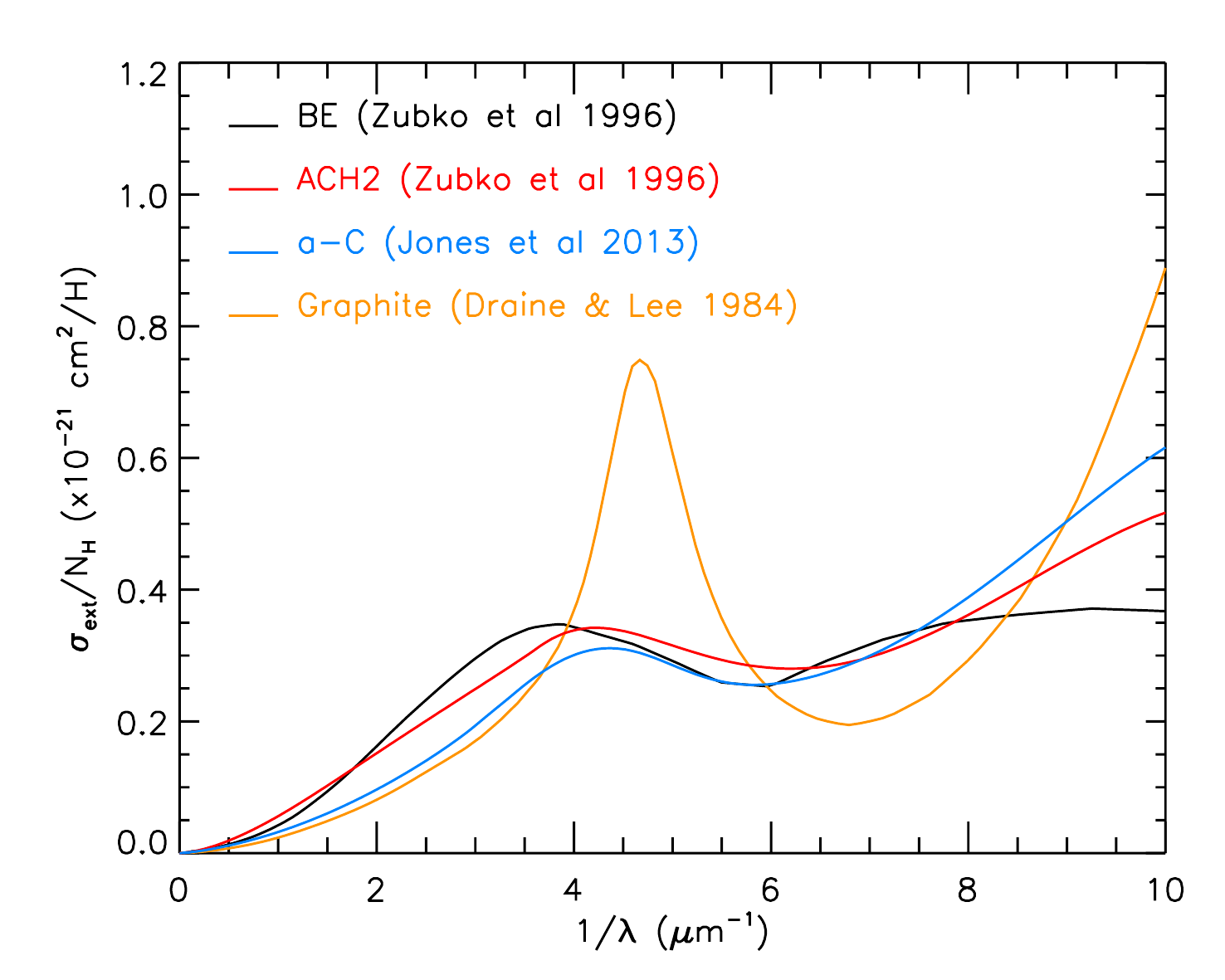}
\caption{Extinction cross-section per H as a function of inverse wavelength, for 0.1\% of the gas mass in the form of carbon, distributed in a power-law from 3 to 100 nm with an index $\alpha = -4.5$. Four optical properties for carbon are compared: BE and ACH2 \citep{Zu96}, a-C(:H) \citep{J13} and graphite \citep{DL84}.}
\label{ext_BE}
\end{figure}

The absorption coefficient of very small carbon grains must present an increase in the UV characteristical for the carbon $\sigma$ to $\sigma^{*}$ transition at 13 eV \citep{J12b}.
This is not the case for the BE sample from \cite{Zu96}. 

Fig.~\ref{ext_BE} compares the extinction curves for 4 different types of carbon grains: the BE and ACH2 samples from \cite{Zu96}, graphite \citep{DL84} and the amorphous carbon a-C(:H) from \cite{J13}. Extinction curves were calculated for a 0.1\% of gas mass in dust, and a power-law distribution from 3 and 100 nm with an index $\alpha = -4.5$ similar to the ones used in the present article for carbon grains. The extinction curves differ one from another: ACH2 and BE samples absorb more than graphite and a-C(:H) in the optical; only graphite present the $\pi$ to $\pi^*$ transition at 4.6 $\mu$m$^{-1}$. Still, all samples except BE present an increase in the UV. 

Because of this unconsistency, the BE sample is probably not the best carbon sample to be used for small carbon grains. This is however of secondary importance in the present article dedicated to the polarized and unpolarized emission of dust grains, dominated by large grains. 
Work is in progress to amend the \cite{J13} dust model for polarization, which will not suffer from this inconsistency.

%
%
\section{Cross-sections for perfect spinning alignment with an inclined magnetic field}\label{MagInclined}
\begin{figure}
\includegraphics[width=\hhsize]{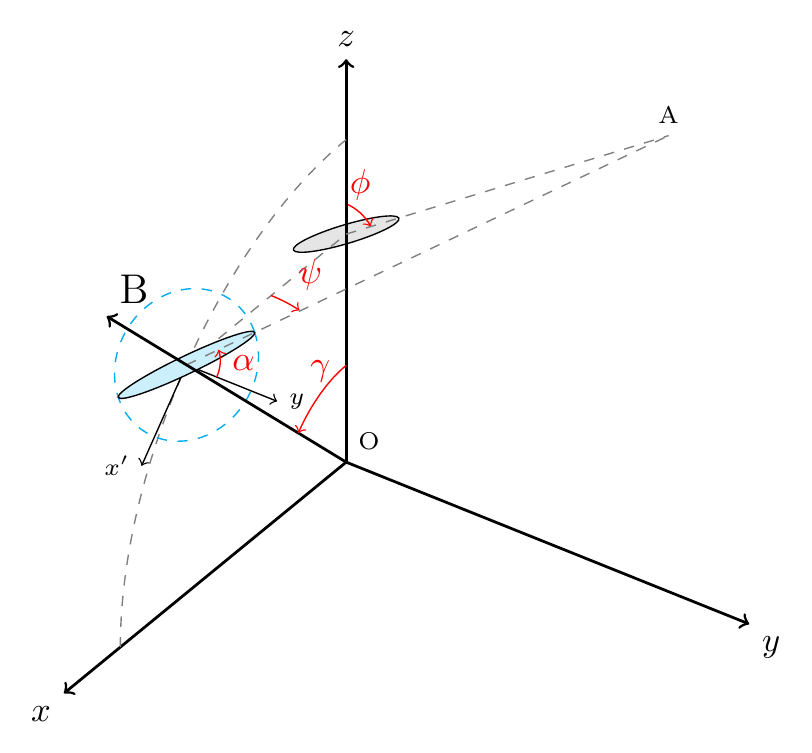}
\caption{Geometry of the problem for a prolate grain (here in cyan) spinning around \Bmag\ in perfect alignment. The axis $\xaxis$ is directed toward the observer. The magnetic field is inclined by an angle $\gamma$ with respect to the plane of the sky $(O,\yaxis,\zaxis)$. $\alpha$ is the spin angle of the grain around \Bmag, measured with respect to $\yaxis$. The projection of the grain onto the plane of the sky is represented in gray. Point $A$ is the intersection of the grain symmetry axis with the plane of the sky. The projected symmetry axis of the grain makes an angle $\phi$ with $\zaxis$. The inclination of the grain with respect to the line of sight $\xaxis$ is $\psi$.}
\label{dynamic}
\end{figure}

Let us now consider the general case where the magnetic field is inclined by an angle $\gamma$ with respect to the plane of the sky, and grains are still in perfect spinning alignment. The cross-section for oblate grains are simply those corresponding to an inclination angle $\incl = \pi/2 - \gamma$. For prolate grains, we must define the geometry of the problem (Fig.~\ref{dynamic}).
In the laboratory frame ($O,\xaxis,\yaxis,\zaxis$), $\xaxis$ is directed toward the observer and 
\Bmag\ lies in the $(O,\xaxis,\zaxis)$ plane making an angle $\gamma$ with $\zaxis$.
The grain is spinning around \Bmag. We note $\alpha$ the spin angle between the symmetry axis of the grain and the $\yaxis$ axis.
The orientation of the symmetry axis $\symaxis$ of the grain with respect to the observer can be obtained with two successive Euler transformations:
\begin{enumerate}
\item spinning: $\symaxis$ is rotated by an angle $\alpha$ around $\zaxis$;
\item inclination: $\symaxis$ is rotated by the angle $\gamma$ around $\yaxis$ to bring the grain spin axis parallel to \Bmag.
\end{enumerate}
After these two rotations, the coordinates of $\symaxis$ are $a_x=-\sin{\alpha}\,\cos{\gamma}$, $a_y=\cos{\alpha}$, and  $a_z=\sin{\alpha}\,\sin{\gamma}$.
The projection of $\symaxis$ onto the plane of the sky makes an angle $\phi$ with $\zaxis$:
\begin{equation}
\tan{\phi} \equiv -\frac{a_y}{a_z} = \frac{-1}{\sin{\gamma}\,\tan{\alpha}} 
\end{equation}
where we follow the IAU sign convention.
The grain symmetry axis $\symaxis$ makes an angle $\incl$ with the line of sight:
\begin{equation}
\cos{\incl} \equiv - a_x = \sin{\alpha}\,\cos{\gamma} 
\end{equation}

Integrating over $\alpha$ on the spinning dynamics of the grain, we find the time-averaged cross-sections \citep{HG80}:
\begin{eqnarray}\label{Eq-CPSAgamma}
C_1 & = & \frac{2}{\pi}\int_0^{\pi/2} \left(\CE\left(\incl\right)\cos^2{\phi} + \CH\left(\incl\right)\sin^2{\phi}\right)\,\d\alpha \\
C_2 & = & \frac{2}{\pi}\int_0^{\pi/2} \left(\CE\left(\incl\right)\sin^2{\phi} +\CH\left(\incl\right)\cos^2{\phi}\right)\,\d\alpha \\
\end{eqnarray}

\begin{acknowledgements}

The research leading to these results has received funding from the European Research Council under the European Union's Seventh Framework Programme (FP7/2007-2013) / ERC grant agreement No. 267934.
V. Guillet thanks the anonymous referee for helpful remarks, and N. Voshchinnikov, R. Siebenmorgen, B. Hensley,  P. G. Martin and G. Green for useful discussions.
This research has made use of NASA's Astrophysics Data System. 

\end{acknowledgements}

\bibliographystyle{aa}
\bibliography{biblio,Planck_bib}

\end{document}